\documentclass[aps,prx,twocolumn,notitlepage,amsfonts,citeautoscript,superscriptaddress,showpacs]{revtex4-1}
\usepackage{graphicx}
\usepackage{amsmath}
\usepackage{physics}
\usepackage{gensymb}
\usepackage{float}
\usepackage{xcolor}
\usepackage{hyperref}
\newcommand{\zl}[1]{\textcolor{black}{#1}}

\newcommand{\LPENS}{\affiliation{Laboratoire de Physique de l'Ecole Normale Supérieure, ENS-PSL, CNRS, Sorbonne Université, Université Paris Cité, Centre Automatique et Systèmes, Mines Paris, Université PSL, Inria, Paris, France}}

\newcommand{\phiext}{\varphi_\mathrm{ext}}

\newcommand{\LPTHE}{\affiliation{Laboratoire de Physique Th\'{e}orique et Hautes Energies, Sorbonne Universit\'{e} and CNRS UMR 7589, 4 place Jussieu, 75252 Paris Cedex 05, France}}

\newcommand{\Google}{\affiliation{Google Quantum AI, Santa Barbara, CA 93117, USA.}}
\newcommand{\Waterloo}{\affiliation{
University of Waterloo, Waterloo, Ontario N2L 3G1, Canada.}}

\date{\today}
\begin{document}

\title{Experimental realization of a $\cos(2\varphi)$ transmon qubit}
\author{E.~Roverc'h}
\thanks{These authors contributed equally to this work}
\LPENS
\author{A.~Borgognoni}
\thanks{These authors contributed equally to this work}
\LPENS
\author{M.~Villiers}
\thanks{These authors contributed equally to this work}
\LPENS
\author{K.~ Gerashchenko}
\LPENS
\author{W.\,C.~Smith}
\Google
\author{C.~Wilson}
\Waterloo
\author{B.~Dou\c{c}ot}
\LPTHE
\author{A.~Petrescu}
\LPENS
\author{P.~Campagne-Ibarcq}
\LPENS
\author{Z.~Leghtas}
\email[]{zaki.leghtas@phys.ens.fr}
\LPENS

\begin{abstract}
    Symmetry-based selection rules make some atomic states remarkably long-lived, a protection superconducting circuits usually lack. We report coherent control and single-shot readout of a 13.6\,MHz $\cos 2\varphi$ qubit in the soft-transmon regime. It consists of an island shunted to ground through an interferometer of Josephson junctions and superinductances that favors tunneling of pairs of Cooper pairs, so that Cooper-pair parity symmetry protects the qubit from dielectric loss. We measure a qubit charge matrix element over 100 times smaller than that of the circuit’s unprotected plasmon transition, placing the dielectric limit on the qubit lifetime above 10\,ms. The measured $T_1 = 70\,\mu$s is instead predicted, with no free parameters, by the $1/f$ flux noise in the interferometer loop that limits dephasing ($T_{2\mathrm{E}} = 2.5\,\mu$s). This experiment demonstrates that Cooper-pair parity symmetry protects against the dielectric loss it targets, and that the remaining limit is the interferometer’s sensitivity to flux noise, which larger superinductances are predicted to reduce without compromising protection.
\end{abstract}

\maketitle

\begin{figure*}[ht]
    \centering
    \includegraphics[scale=1]{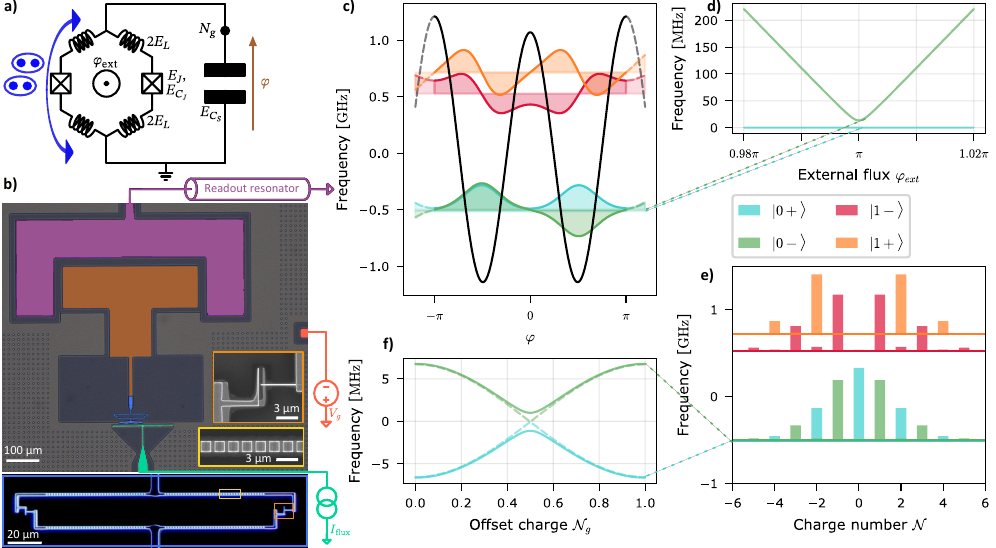}
    \caption{{\bf Device implementation.} (a) Lumped-element circuit of a capacitor shunted by a KITE loop that transfers pairs of Cooper pairs (blue). (b) Optical false-color micrograph of the device. Pairs of Cooper pairs tunnel in and out of the central island (brown) through the KITE (dark-field optical micrograph in blue inset) composed of chains of JJs (SEM image in yellow inset) and small JJs (SEM image in orange inset). A flux line (green) threads magnetic flux through the KITE loop, while a charge line  (red) carries radio-frequency drives and controls the offset charge. A quarter wave-length resonator (purple) is capacitively coupled to the island for dispersive readout. While the KITE is in aluminium, all other metallic components are in niobium on silicon. (c) Four lowest energy levels and wavefunctions confined in the double-well potential (black line) of Hamiltonian \eqref{eq:H1M} at $\mathcal{N}_g=0, \phiext=\pi$. (d) Flux dispersion of the two lowest energy levels of Hamiltonian \eqref{eq:H1M} at $\mathcal{N}_g=0$. (e) Same as panel (c) in the discrete charge representation $\mathcal{N}$. (f) Charge dispersion of the ground state doublet at $\phiext=\pi$. The avoided crossing at $\mathcal{N}_g=0.5$ stems from the KITE asymmetry. The case of perfect symmetry is overlaid in dashed lines for clarity.}
    \label{fig:fig1}
\end{figure*}
\section{Introduction}
Quantum technology platforms continue to battle against decoherence for the emergence of large scale quantum computers. In superconducting circuits, a widespread strategy is to employ a simple physical qubit (namely the transmon \cite{Koch2007}), and improve coherence at the logical level through active quantum error correction \cite{Krinner2022,Google2025quantum}. The main challenge is the large qubit overhead required to reach computationally relevant error rates \cite{Gidney2025}. An alternative and complementary route is to engineer more complex physical qubits with intrinsically reduced error channels, thereby lowering the overhead of outer error-correction layers \cite{Ruiz2025ldpc, Gouzien2023}. Within this paradigm, two broad classes of qubits have emerged. Bosonic qubits exploit relatively simple circuit Hamiltonians, involving low-order nonlinearities such as Kerr interactions or two-photon exchange, and encode information in states stabilized by tailored drives or parametric pumping \cite{Joshi2021}. Protected qubits, by contrast, rely on static circuit Hamiltonians formed from complex arrangements of superconducting elements supporting multiple strongly coupled modes \cite{GyenisReview}. The circuit is engineered such that two levels separate from the rest of the spectrum and display a grid-like structure, showcasing non-overlapping supports and delocalization across \zl{coordinate}-space \cite{GKP2001}. Some qubit proposals lie at the boundary of these families, where both strong nonlinearities and parametric pumps coexist \cite{Wang2024kapitza,Sellem2025}.

Symmetries lie at the core of qubit protection \cite{Doucot2012, Lin2018}. The circuit Hamiltonian displays a symmetry, with the qubit states residing in distinct symmetry sectors of the spectrum. Prominent examples include translational symmetry in GKP qubits \cite{GKP2001,Campagne2020}, fluxon parity in the bifluxon \cite{Kalashnikov2020} or the integer fluxonium \cite{Ardati2024, Mencia2024}, and Cooper-pair number parity in the $\cos(2\varphi)$ qubit \cite{Doucot2002, Ioffe2002}. \zl{Beyond their prospective role as qubits, circuits engineered to bound one or more noise channels by construction also serve as instruments of discovery: any measured coherence below the predicted bound signals an unidentified loss mechanism \cite{pop2014,Reglade2024}.}

The $\cos(2\varphi)$ qubit, central to this work, consists of a superconducting island of charging energy $\mathcal{E}_C$ shunted to ground by a tunneling element of energy scale $\mathcal{E}_{J2}$ that permits only pairs of Cooper-pairs to tunnel; Cooper-quartet tunneling translates into a $\mathcal{E}_{J2}\cos(2\varphi)$ potential in the phase $\varphi$ conjugate to charge $\mathcal{N}$. This circuit is described at offset charge $\mathcal{N}_g$ by Hamiltonian
\begin{equation}
    \mathcal{H}_{\mathrm{cos}2\varphi}=4\mathcal{E}_C (\mathcal{\hat N}-\mathcal{N}_g)^2+\mathcal{E}_{J2}\cos(2\hat \varphi)\;.
    \label{eq:Hcos2phi}
\end{equation}
Proposals and realizations of Cooper-quartet tunneling include d-wave superconductor junctions \cite{Brosco2024flowermon}, gate-tunable weak links \cite{Larsen2020,Messelot2024,Leblanc2025}, and,  addressed hereafter, two-identical-arm interferometers composed of multiple Josephson junctions (JJ) and biased at half of a flux quantum \cite{Gladchenko2008, Bell2014, Smith2020, Dodge2023, Hays2025, Sanchez2026, Zhurbina2026} (Fig.~\ref{fig:fig1}a).

The experimental implementation of a $\cos(2\varphi)$ qubit was pioneered in Ref \cite{Bell2014}. The circuit was operated in a light regime of $\mathcal{E}_{J2}/\mathcal{E}_C\approx 4$, leading to a broad 4 GHz charge dispersion, akin to a Cooper-pair box regime. Recent experiments have delved deeper in the transmon regime \cite{Larsen2020, Gyenis0pi, Nguyen2025, feldsteinbofill2026}, but operated away from the symmetry point or only demonstrated coherent manipulations between higher plasmonic transitions. Unlike the transmon states that correspond to plasmonic excitations of a Josephson well, the $\cos(2\varphi)$ lowest doublet states correspond to bonding and antibonding combinations of fluxon states (Fig.\ref{fig:fig1}c). The \zl{doublet} energy splitting, which is a tunneling energy, is therefore equal to the charge dispersion \cite{Koch2007} (Fig.\ref{fig:fig1}f). Ideally, diving deep in the transmon regime eliminates sensitivity to charge, but results in a vanishingly small qubit frequency that exacerbates sensitivity to flux \cite{messelot2026} and complicates measurement. A better compromise is the so-called ``soft'' \cite{GyenisReview} transmon regime, where the qubit frequency is small enough so that charge noise is non-limiting, but control and readout are possible.


In this experiment, we implement a $\cos(2\varphi)$ qubit operating in the soft transmon regime with $\mathcal{E}_{J2}/\mathcal{E}_C=22$. The qubit doublet lies at 13.6~MHz, 400 times smaller than Ref.~\cite{Bell2014}, which pushes charge-induced errors well above the measured coherence times. Despite this low frequency, we demonstrate coherent control and single-shot readout. This was made possible by careful design choices, detailed below, of our tunneling element coined the ``KITE'' \cite{Smith2020, Smith2022, Smith2025} (Kinetic Interference co-Tunneling Element). We achieve $83~\%$ single-shot fidelity in an integration time of $5~\mu$s, thereby initializing the doublet at an effective temperature of $400~\mu$K. The protection of the qubit charge degree of freedom is measured through a 100-fold reduction in the charge matrix element compared to an unprotected plasmon transition. This sets an estimated bound on dielectric-induced decay of 10~ms, limited by asymmetry in the KITE. Instead, we measure $T_1 = 70\zl{\pm8}~\mu\mathrm{s}, T^\mathrm{echo}_2 =2.5 \zl{\pm 0.1}~\mu\mathrm{s}$, limited by $1/f$ flux noise in the KITE loop \cite{Yan2016}. \zl{We then identify KITE parameters that reduce flux-sensitivity by two orders of magnitude without compromising dielectric loss protection. Combining this new regime with smaller loop areas, gradiometric designs or fluxoid locking \cite{Majer2002,Schwarz2013, Benatre2025, Hida2025}, could push flux-noise-limited coherence to the millisecond range.}

\section{Circuit implementation}

Our device (Fig.~\ref{fig:fig1}b) is fabricated in a 2D coplanar waveguide geometry, anchored at the 10 mK plate of a dilution refrigerator, and measured in a standard cQED setup (\ref{sec:samplesetup}). It consists of a large $0.2~\mathrm{mm}^2$ niobium island of charging energy $E_{C_S}$, capacitively coupled from the right to a charge bias line, from the top to a readout resonator, and shunted to ground through a KITE. Each arm of the KITE contains a small-area junction of Josephson and charging energies $E_J$ and $E_{C_J}$, in series with a superinductance of inductive energy $E_L$ made of an array of 150 larger-area junctions. An on-chip flux-bias line threads flux $\varphi_\mathrm{ext}$ into the KITE loop. The readout mode is a quarter-waveguide resonator at $4.3~\mathrm{GHz}$, inductively coupled at rate $\kappa/2\pi\approx700~\mathrm{kHz}$ to a transmission line connected to a traveling wave parametric amplifier (TWPA).

Quantizing the circuit of Fig.~\ref{fig:fig1}a leads to a three-mode Hamiltonian (\ref{sec:fit}). Near $\phiext=\pi$, our device's low-energy dynamics reduce to an effective 1-mode Hamiltonian:
\begin{equation}
\mathcal{H}_\mathrm{circuit}^\text{1-mode}=\mathcal{H}_{\mathrm{cos}2\varphi}-\mathcal{E}_{J1}\cos(\hat\varphi)-\mathcal{E}_{J\phi}\sin(\hat\varphi)(\phiext-\pi)\;.
    \label{eq:H1M}
\end{equation}
At $\phiext=\pi$, this corresponds to an ideal $\cos(2\varphi)$ Hamiltonian with an additional single Cooper-pair tunneling amplitude $\mathcal{E}_{J1}$ due to KITE asymmetry. The last term in Eq.~\eqref{eq:H1M} accounts for the sensitivity of the device to flux, and lifts the degeneracy of the two potential wells away from $\pi$.

Let us now explain how we choose our circuit parameters and how they map to Hamiltonian~\eqref{eq:H1M}. While we require the $\cos(2\varphi)$ transmon mode to be heavy, the internal loop modes are engineered to be light, such that they self-resonate at high frequencies and merely shape the effective $\mathcal{E}_{J2}\cos(2\varphi)$ potential. This design requirement leads to $E_{C_S}\ll E_{C_J}$. We find that a sizable $\mathcal{E}_{J2}$ requires $E_J\gtrsim E_{C_J}$, which in turn sets $\mathcal{E}_{J2}\approx E_L$ \cite{Roverchtheory}. Since $\mathcal{E}_C\approx E_{C_S}$, the transmon regime then demands $E_L\gg E_{C_S}$. Finally, akin to the evolution from the flux qubit to the fluxonium \cite{manucharyan2009fluxonium}, the sensitivity to flux through the $\mathcal{E}_{J\phi}$ term is softened by imposing $E_L\ll E_J$, which in turn sets $\mathcal{E}_{J\phi} \approx \frac{\pi}{2}E_L$.  The set of circuit parameters we chose to satisfy these constraints and the corresponding $\cos(2\varphi)$ parameters are displayed in Table~\ref{table1}. A detailed mathematical analysis of this correspondence, based on the Born-Oppenheimer approximation \cite{BO1927},  is derived in Ref.~\cite{Roverchtheory}. 

\begin{table}[h]
\begin{ruledtabular}
\begin{tabular}{ccccc}
 \multicolumn{5}{c}{Lumped-element circuit parameters} \\\cline{1-5}
$E_J \text{[GHz]}$ &$E_{C_J} \text{[GHz]}$& $E_L \text{[GHz]}$ &$E_{C_S} \text{[MHz]}$& $\varepsilon$ \\
\hline
16.83 & $4.82$ & $1.27$ & $72$ & $3\%$ \\

\end{tabular}
\begin{tabular}{cccc}
 \multicolumn{4}{c}{$\cos(2\varphi)$ transmon model} \\\cline{1-4}
$\mathcal{E}_{J2} \text{[GHz]}$ &$\mathcal{E}_{J1}\text{[GHz]}$ & $\mathcal{E}_{J\phi}\text{[GHz]}$ &$\mathcal{E}_{C} \text{[MHz]}$ \\
\hline
$1.14$ & $0.07$ & $1.92$&$52$ \\

\end{tabular}
\end{ruledtabular}
\caption{Parameters of the circuit elements of Fig.~\ref{fig:fig1}a and the corresponding quantities entering the single-mode Hamiltonian~\eqref{eq:H1M}. The parameter $\varepsilon$ corresponds to the relative  asymmetry of the two small junctions.}
\label{table1}
\end{table}
Two types of transitions appear in the spectrum of Hamiltonian \eqref{eq:H1M} (Fig.~\ref{fig:fig1}c,e). Intrawell transitions, referred to as plasmons, are quasiharmonic and disperse weakly with flux. In addition, away from $\pi$ flux, plasmons disperse weakly with charge and resonate at frequency $\omega_\mathrm{plasmon} \approx \sqrt{32 \mathcal{E}_{J2}\mathcal{E}_{C}}/\hbar$ \cite{Smith2020}. Conversely, interwell transitions, referred to as fluxons, disperse strongly with flux (Fig.~\ref{fig:fig1}d), with doublets forming at the half flux quantum sweet spot. Energy levels are labeled $\ket{n\pm}$ with $n$ and $\pm$ denoting the plasmon and Cooper-pair-parity quantum numbers respectively.

We find inspiration in the heavy fluxonium design \cite{ Zhang2021} to obtain a high-fidelity readout of the low-frequency doublet. Starting from either of the two doublet states, the corresponding ladder of plasmonic transitions induces a frequency pull on the readout mode \cite{Zhu2013}. The readout coupling strength and frequency are designed to obtain a large differential frequency pull depending on the doublet state. This direct qubit readout relies on operation in the soft transmon regime; indeed, delving deeper in the transmon limit (i.e increasing $\mathcal{E}_{J2}/\mathcal{E}_{C}$), the two ladders would become indistinguishable to the readout mode.

\section{Readout}

\begin{figure}[h!]
    \centering
    \includegraphics[scale=1]{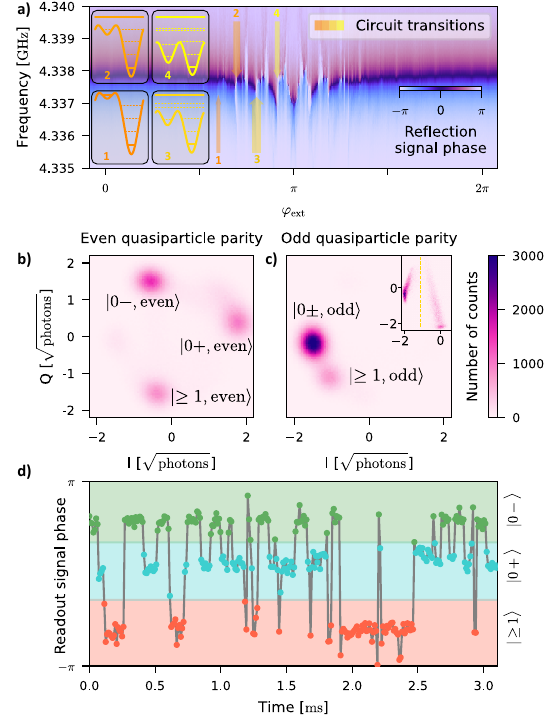}
    \caption{ {\bf Readout mode spectroscopy and single shot measurements}. (a) Reflected signal phase (color) versus probe frequency (y-axis) and external flux (x-axis). Anti-crossings (arrows) appear when a circuit transition (insets) collides with the readout frequency. The regions of frequency collision are broadened by charge dispersion, and are encoded in the arrow widths. (b,c) Two-dimensional histograms of the measured $(I,Q)$ values at resonance, $N_g=0$ and $\phiext=\pi$, integrated over $10~\mu\mathrm{s}$ after post-selection on even (b) or odd (c) quasiparticle number.  \zl{The inset of panel (c) shows the quasiparticle-selection measurement histogram, acquired from measurements integrated over $200~\mu\mathrm{s}$, where two distributions corresponding to either quasiparticle parity appear (see \ref{sec:quasiparticles}).} Labels indicate the state associated to each distribution. (d) Measured readout-signal phase (y-axis) versus time (x-axis), after post-selection on even quasiparticle number. The telegraphic signal transits between three zones corresponding to $\ket{0-}$ (green), $\ket{0+}$ (cyan), and higher plasmon states denoted $\ket{\ge 1}$ (orange).}
    \label{fig:fig2}
\end{figure}

We characterize our readout mode in Fig.~\ref{fig:fig2}. We start by acquiring the readout reflection spectrum versus external flux. At $\phiext=0$, the readout response is a canonical $2\pi$ phase roll. Conversely, around $\phiext=\pi$, multiple features are visible. Anticrossings are expected when a fluxon transition, accompanied with $n$ plasmonic excitations, comes into resonance with the readout mode: $\omega_\mathrm{readout}\approx2 \mathcal{E}_{J\phi}\abs{\varphi_\mathrm{ext}-\pi}/\hbar+n\times\omega_\mathrm{plasmon}$. In our system where $\omega_\mathrm{readout}\lesssim 4 \times \omega_\mathrm{plasmon}$, exactly 4 anticrossings per half period are expected, corresponding to $n=0,1,2,3$. In this device, more features are visible, possibly due to higher-order transitions and transitions from thermally-occupied states. Some of these features are sharp, and are therefore used for flux calibration (\ref{sec:biascal}).

The circuit contains a superconducting island and is therefore sensitive to charge. Charge fluctuations arise from both continuous electrostatic variations, due to the gate voltage and charge motion in the substrate, and discrete $1e$ jumps caused by quasiparticle poisoning \cite{Catelani2011}. From the readout response versus gate voltage, we calibrate the bias in units of offset charge $\mathcal{N}_g$, and correct for slow electrostatic drifts (\ref{sec:biascal}). Moreover, we acquire histograms of measurement shots integrated over $200~\mu\mathrm{s}$, a time scale that is large enough to average out qubit dynamics, but short enough to resolve quasiparticle tunneling events. Two distributions appear, corresponding to even and odd quasiparticle numbers on the island with a characteristic switching time of $56~\mathrm{ms}$ (\ref{sec:quasiparticles}).

We now set $\phiext=\pi$ and $\mathcal{N}_g=0$. Since the readout mode is expected to display qubit-state-dependent resonance frequencies, the phase of a measurement pulse reflecting off the readout mode encodes the qubit state. We acquire histograms of measurement shots with an integration window $T_\mathrm{int} = 10~\mu\mathrm{s}$. We observe well-separated distributions for both quasiparticle parities. 
The readout power (estimated to $\bar{n}\approx2.3$ circulating photons, see \ref{sec:photonnumber}), and frequency are adjusted to maximize the separation between the $\ket{0+}$ and $\ket{0-}$ distributions at even quasiparticle number. In summary, this experiment enables direct single-shot readout of quasiparticle parity, plasmonic excitation, and Cooper-pair parity within the ground state doublet.

We demonstrate the quantum non-demolition (QND) character of our high-fidelity measurement in Fig.~\ref{fig:fig2}(d). A sequence of back-to-back $10~\mu\mathrm{s}$ measurement pulses is applied to the readout mode, with an average intraresonator photon number $\bar{n}=2.3$. For each pulse, we extract an $(I,Q)$ pair and plot the phase of $I+iQ$ as a function of time. The resulting signal exhibits a telegraph behavior, characteristic of quantum jumps. Real-time monitoring reveals transitions of the circuit between $\ket{0+}$, $\ket{0-}$, and higher-energy levels. Fitting the average occupation of each sector to the thermal population of a harmonic ladder at the plasmon frequency yields a circuit temperature $T=43~\mathrm{mK}$. Remarkably, we are able to resolve single Cooper-pair tunneling events, even though the associated energy-level splitting amounts to only $0.3~\%$ of the readout-mode frequency.

\begin{figure}[h!]
    \centering
    \includegraphics[scale=1]{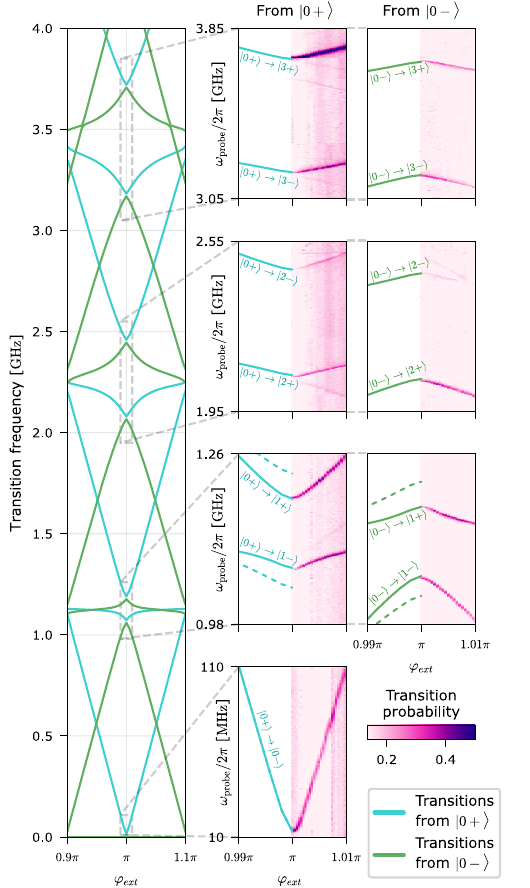}
    \caption{{\bf Device spectroscopy}. (left) Transition frequencies (y-axis) from $\ket{0+}$ (cyan) and $\ket{0-}$ (green) versus external flux (x-axis), calculated from the three-mode Hamiltonian (\ref{sec:fit}) at $N_g=0$. (center, right) Two-tone spectroscopy data around the first 4 plasmon doublets in the vicinity of $\pi$. We measure the transition probability out of the ground state doublet (color) from $\ket{0+}$ (center) and $\ket{0-}$ (right) versus external flux (x-axis) and the frequency of the probe tone (y-axis). For the ground state doublet (lowest frame), we instead represent the transition probability from $\ket{0+}$ to $\ket{0-}$. The three-mode model (solid lines) reproduces the spectra over the full measured bandwidth. The one-mode model (dashed lines), valid at low energies, captures the two lowest doublets with a $3\%$ error. Theory curves are displayed \zl{on the left half of the panels and data on the right} for clarity.}
    \label{fig:fig3}
\end{figure}

\section{Spectroscopy}

We further characterize our circuit in Fig.~\ref{fig:fig3} through two-tone spectroscopy versus external flux at $\mathcal{N}_g=0$ (see \ref{sec:oddmanifold} for $\mathcal{N}_g=0.5$). At each flux point, we apply a $10~\mu\mathrm{s}$ probe tone through the charge line of variable frequency in the 10~MHz to 4~GHz range. This pulse is preceded and followed by a single-shot $5~\mu\mathrm{s}$ readout pulse with $\bar{n}=4$ circulating photons. By analyzing the correlations between the two measurements, we extract the probability of population transfer out of $\ket{0\pm}$ versus the probe tone frequency. Charge and flux calibration protocols are interleaved every 30 and 90 seconds respectively to compensate slow drifts, and a quasiparticle parity measurement is performed every $10~\mathrm{ms}$ to only keep data in the even quasiparticle parity sector.

We fit the data to the full three-mode circuit Hamiltonian (solid lines in Fig.~\ref{fig:fig3}). Transitions originating from both $\ket{0+}$ and $\ket{0-}$ are shown. Plasmon transitions (around $1.1~\mathrm{GHz}$, $2.3~\mathrm{GHz}$, and $3.4~\mathrm{GHz}$) are readily identified by their relatively weak flux dependence and their quasi-harmonic spacing. By contrast, fluxon transitions disperse strongly with flux, exhibiting a linear dependence with slope $\approx \pi E_L$ away from the sweet spot.

Remarkably, the broadband spectrum of our circuit containing 302 JJs, one loop, and one island is well captured by the 3-mode model with five fit parameters. Moreover, the two lowest doublets are also well captured by the 1-mode model described by Hamiltonian~\eqref{eq:H1M} (dashed lines in Fig.~\ref{fig:fig3}) with 4 fit parameters, and as expected, the agreement with this low-energy model worsens as we climb the plasmonic ladder. The extracted parameters are summarized in Table \ref{table1}.


\section{Qubit control}
In principle, a qubit protected from charge-induced noise should also be insensitive to control drives applied through the charge line. One alternative would be to control the qubit through the unprotected flux channel. Here, for ease of operation, we apply all control signals through the charge line. Driving the protected qubit transition remains possible due to the finite asymmetry of the KITE element (here $3\%$), which provides residual coupling to the charge line. We compensate the resulting weak coupling with increased input power, while leakage to higher excited states remains small due to the large spectral gap ($\approx 1~\mathrm{GHz}$).

\begin{figure}[h!]
    \centering
    \includegraphics[scale=1]{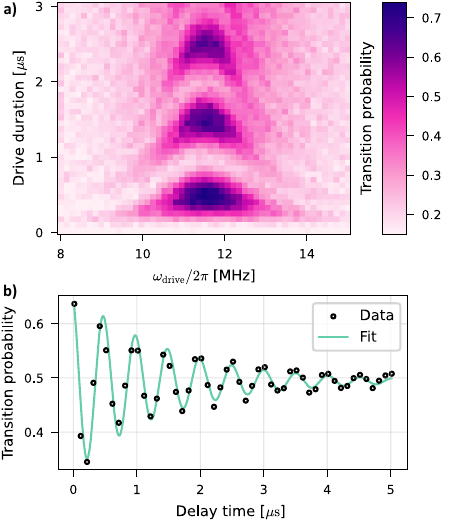}
    \caption{{\bf Rabi chevrons and Ramsey spectroscopy}. (a) Measured transition probability (color) from $\ket{0+}$ to $\ket{0-}$ in the presence of a Rabi drive of varying duration (y-axis) and frequency (x-axis). (b) Transition probability from $\ket{0+}$ to $\ket{0-}$ (y-axis) versus delay time (x-axis) between the two $\pi/2$ pulses of the Ramsey sequence. The data (open circles) fit an exponentially decaying envelope with a Ramsey decay time $T_2^\mathrm{Ramsey}=1.44~\mu\mathrm{s}$. The signal oscillations arise from the combined digital and physical detunings of the $\pi/2$ pulses with respect to the bare qubit frequency.}
    \label{fig:fig4}
\end{figure}

Coherent control of our $\cos(2\varphi)$ qubit is shown in Fig.~\ref{fig:fig4}(a). We set $\mathcal{N}_g=0$ and $\phiext=\pi$, and apply a drive pulse through the charge line of variable duration and frequency. This pulse is preceded and followed by a readout pulse, and calibration routines are interleaved as previously explained. We observe canonical Rabi chevrons, demonstrating coherent oscillations between states of opposite Cooper-pair parity. The shift between the chevron symmetry point (11.5 MHz) and the bare frequency measured through a Ramsey sequence $\omega_q/2\pi = 13.6\  \mathrm{ MHz}$  (Fig.~\ref{fig:fig4}b) is consistent with an AC Stark shift of the doublet due to off-resonant driving of plasmonic states \cite{Gerashchenko2025}. At resonance, the achieved Rabi rate $\Omega_\mathrm{Rabi} = 2\pi\times 1~\mathrm{MHz}$ is $\approx 10\%$ of the transition frequency, placing the system at the boundary of the regime where the rotating-wave approximation is valid. A notable feature of our qubit is the large separation ($\approx 1~\mathrm{GHz}$) from other circuit transitions, a property that could be leveraged for fast and high-fidelity gates \cite{Zhang2021}. The oscillations decay at $T_2^\mathrm{Rabi}= 4 \zl{\pm0.3}~\mu\mathrm{s}$, exceeding the measured Ramsey time $T_2^\mathrm{Ramsey}= 1.4\zl{\pm0.1}~\mu\mathrm{s}$, and therefore hinting towards decoherence dominated by low-frequency noise.


\section{Decoherence analysis}
\label{sec:decoherence}

\begin{figure}
    \centering
    \includegraphics[scale=1]{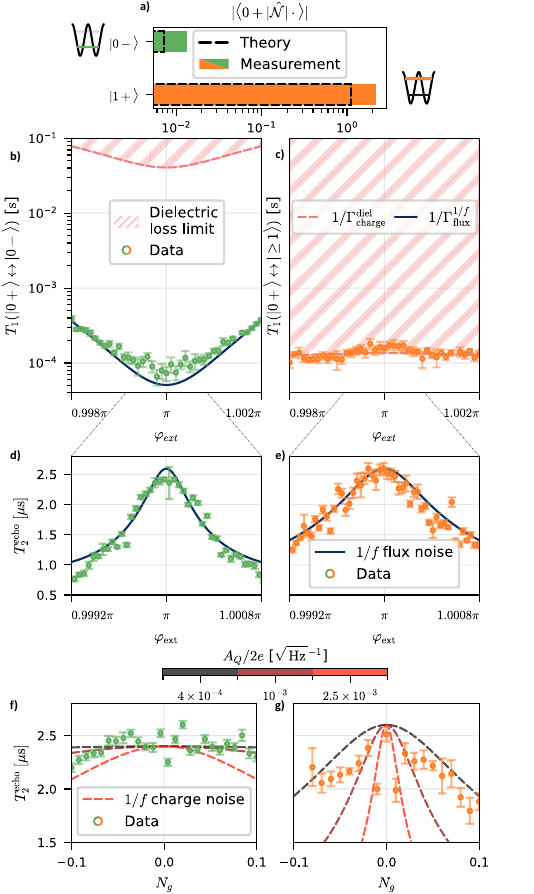}
    \caption{{\bf Analysis of device losses}. (a) Island charge matrix element (x-axis) for the doublet and plasmon transitions (y-axis). Theoretical predictions from Eq.~\eqref{eq:H1M} (dashed bars) agree with the measurements (colored bars), and remaining discrepancies are attributed to uncertainty in the attenuation of the input lines. (b,c) Decay time (y-axis) of the doublet (b) and the plasmon (c) versus external flux (x-axis). The data (circles) are compared to calculated contributions from Eqs.~\eqref{eq:diel},\eqref{eq:1of} (solid lines) with the relevant end states and associated transition frequencies. The dielectric-induced loss limit (red hatched zone) is two orders of magnitude larger for the protected doublet than the unprotected plasmon. (d,e) Echo decay time (y-axis) versus external flux (x-axis) for the doublet (d) and plasmon (e) transitions. We fit the data (circles) to a $1/f$ flux noise model (solid line). (f,g) Echo decay time (y-axis) versus offset charge (x-axis) for the doublet (f) and plasmon (g) transitions. We fit the data (circles) to a $1/f$ charge noise model (dashed lines) for three values of the noise amplitude (color). Error bars are estimated from a bootstrap method (\ref{sec:t1}).}
    \label{fig:fig5}
\end{figure}

The underlying premise of the $\cos(2\varphi)$ qubit is to protect against dielectric loss by encoding information in states with non-overlapping supports in the charge basis. The associated loss rate reads  
\begin{equation}
    \Gamma^\mathrm{diel}_\mathrm{charge}=\frac{16\mathcal{E}_C}{\hbar Q_\mathrm{cap}}\coth(\frac{\hbar\omega_q}{2k_B T})\abs{\bra{0+}\mathcal{\hat N}\ket{0-}}^2 ,\label{eq:diel}
\end{equation}
where $1/Q_\mathrm{cap}$ is the dielectric loss tangent, $k_B$ is Boltzmann's constant and $\hbar$ is the reduced Planck constant. \zl{While in a transmon qubit the charge matrix element is of order unity, here it vanishes proportionally to KITE asymmetry}. We test this protection by directly comparing the island charge matrix elements of the protected $\ket{0+}\leftrightarrow \ket{0-}$ fluxon transition and an unprotected \(\ket{0+}\!\leftrightarrow\!\ket{1+}\) plasmon transition (Fig.~\ref{fig:fig5}a). To this end, we measure the microwave power required for a $\pi$ pulse through the charge line for each transition; the independently calibrated line attenuation then yields the corresponding charge matrix elements. We find a charge matrix element of $2.2$ for the unprotected transition, comparable to that of a conventional transmon, whereas the one for the protected transition is strongly suppressed to $1.3\times10^{-2}$. This suppression is limited by KITE asymmetry and is consistent with theoretical predictions within the uncertainties in line attenuation.

We analyze the contributions of various loss mechanisms to our qubit $T_1$ in Fig.~\ref{fig:fig5}b,c. We operate in the even quasiparticle parity sector, set $\mathcal{N}_g=0$ and $\phiext$ in an interval around $\pi$ and perform two readout measurements separated by a wait time $\tau$. By correlating these two measurements, we extract the transition probabilities between states $\{\ket{0+},\ket{0-},\ket{\ge 1}\}$ as a function of wait time $\tau$, leading to the plotted measured lifetimes (\ref{sec:t1}). The unprotected plasmon transition is limited by dielectric loss, and from its lifetime measurement we extract a lower bound on the dielectric quality factor $Q_\mathrm{cap}\approx  1.5\times 10^6$. This leads to an estimated limit for the doublet lifetime at half-flux of $1/\Gamma^\mathrm{diel}_\mathrm{charge}\approx 40~\mathrm{ms}$, and the limits due to charge noise and radiative loss through the charge line are orders of magnitude higher.

Having established a high dielectric-loss limit, we next consider inductive loss mechanisms. We find that the measured lifetime of the doublet closely matches the one predicted from coupling to a bath of spins following a $1/f$ noise spectral density. Following the conventions of Ref.~\cite{Yan2016}, the associated rate reads
\begin{equation}
    \Gamma^{1/f}_\mathrm{flux}= 2\left(\frac{2\pi\mathcal{E}_{J\phi}}{\Phi_0\hbar}\right)^2 S_{\Phi\Phi}(\omega_q)\abs{\bra{0+} \sin(\hat\varphi) \ket{0-}}^2 \;,\label{eq:1of}
\end{equation}
where $\Phi_0 = \frac{h}{2e}$ is the superconducting flux quantum and the flux noise spectral density is $S_{\Phi\Phi}(\omega_q)=A_\Phi^2\times (2\pi \times 1~\mathrm{Hz})/\omega_q$. The noise amplitude $A_\Phi=\zl{5.25 \pm 0.07}~\mu \Phi_0 / \sqrt{\mathrm{Hz}}$, extracted from $T_2^\mathrm{echo}$ measurements (Fig.~\ref{fig:fig5}d,e), is about 4 times larger than state of the art~\cite{Yan2016}, which is consistent with our large loop dimensions. Remarkably, the theory (dark blue curve in Fig.~\ref{fig:fig5}b) has no free fit parameters, showing that the doublet $T_1$ is limited by flux noise. While the plasmon $T_1$ depends weakly on flux, the doublet $T_1$ rises from \zl{$70\pm 8~\mu\mathrm{s}$} at $\phiext=\pi$ to \zl{$300\pm14~\mu\mathrm{s}$} at $\phiext=\pi\times(1\pm 0.002)$. Indeed, shifting $\phiext$ from $\pi$ breaks the degeneracy of the double-well potential and the wavefunctions associated to $\ket{0\pm}$ each localize in one of the two wells, thereby suppressing their overlap in phase and increasing $T_1$. Finally, the limit due to radiative loss through the flux line is two orders of magnitude larger than the measured lifetimes, which in turn set a lower bound on a hypothetical inductive quality factor associated to the KITE loop to $Q_\mathrm{ind}\ge 5\times 10^8$ (\ref{sec:t1}) \cite{pop2014}.

We characterize the device coherence versus flux through $T_2$ echo measurements (Fig.~\ref{fig:fig5}d,e). Following Ref.~\cite{Zhang2021}, the data are fit to $\exp[-\Gamma_\nu t-(\Gamma_\phi t)^2]$, where $\Gamma_\nu$ is flux-independent and $\Gamma_\phi$ arises from $1/f$ flux noise (\ref{sec:t2}). We observe a rapid reduction of coherence for flux offsets as small a few $10^{-4}~\Phi_0$, indicating that away from the sweet spot coherence is limited by flux noise. \zl{In principle, placing ourselves exactly at $\pi$ would render the device first-order insensitive to flux noise. In practice however, our flux calibration has a standard deviation of $\approx 8\times 10^{-5}\Phi_0$ (extracted from Fig.~\ref{fig:fluxbias}), which qualitatively explains the observed maxima in the data of Fig.~\ref{fig:fig5}d,e}.

In contrast, the echo time shows almost no dependence on charge offset (Fig.~\ref{fig:fig5}f,g). Theory curves for charge-noise-induced decoherence are overlaid on the data for various values of the charge noise amplitude $A_Q$. Although the data trends are not strong enough to conclusively give a value for $A_Q$, we can nevertheless infer an upper bound $A_Q \leq 2\times10^{-3}e/\sqrt{\mathrm{Hz}}$. Based on Ref.~\cite{Smith2019}, we estimate the charge-induced decoherence for the doublet at zero offset charge at $\approx 1~\mathrm{ms}$. A more conservative bound  $\approx 350~\mu\mathrm{s}$ is obtained by extrapolating the measured echo time in Ref.~\cite{Bell2014} to our 400-fold reduced charge dispersion. By operating in the soft transmon regime, charge-induced decoherence has been pushed two orders of magnitude above the observed limit.

\zl{We end this section by emphasizing that each noise parameter entering this analysis is extracted independently from a dedicated measurement. The dielectric quality factor $Q_\mathrm{cap}$ is extracted solely from the unprotected plasmon $T_1$ (Fig.~\ref{fig:fig5}c), and applied without adjustment to predict the doublet's dielectric-loss limit (Fig.~\ref{fig:fig5}b, red hatched region). Similarly, the flux-noise amplitude $A_\Phi$ is extracted from $T_2^\mathrm{echo}$ versus flux (Fig.~\ref{fig:fig5}d,e), independent of the $T_1$ data it is then used to predict (dark blue curve, Fig.~\ref{fig:fig5}b), which has no free parameters. Finally, the central claim of charge-parity protection is supported by a direct measurement of the charge matrix element of the doublet, that is then compared to the unprotected plasmon and model predictions (Fig.~\ref{fig:fig5}a).}

\section{Beyond single-mode analysis}
\label{sec:discussion}
A recurring conundrum in protected qubits is that the circuit complexity needed to realize promising Hamiltonians introduces additional dissipation channels through extra loops, islands, or modes. A notable example is the so-called $\zeta$-mode in the $0-\pi$ qubit \cite{Groszkowski2018}.

In this section, we discuss loss mechanisms that evade our single-mode analysis. As a first refinement, we consider the three-mode model (\ref{sec:fit}) describing the circuit shown in Fig.~\ref{fig:fig1}a. While dielectric loss in the shunt capacitance and $1/f$ flux noise in the KITE loop are well captured by the single-mode description, an additional dissipation channel arises from dielectric loss in the capacitances of the two small Josephson junctions. In the symmetric regime, this loss rate reads (\ref{sec:t1})
\begin{eqnarray*}
    \Gamma_\mathrm{charge}^\mathrm{JJ} &=& \frac{\hbar\omega_q^2}{2E_{C_J} Q_\mathrm{cap}}\coth(\frac{\hbar\omega_q}{2k_B T})\abs{\bra{0+}\hat \phi_\Delta \ket{0-}}^2\\
    &\approx& \frac{k_B T}{Q_\mathrm{cap}}\frac{\omega_q}{E_{C_J} }\abs{\bra{0+}\hat \phi_\Delta \ket{0-}}^2\;,
    \label{eq:JJdiel}
\end{eqnarray*}
where $\hat \phi_\Delta$ is the phase operator associated to the circulating mode in the KITE loop, and the last approximation applies since $\hbar\omega_q\ll k_B T$. This expression follows a form similar to that of dielectric loss in a heavy fluxonium. A key distinction is that it scales with the ratio $\omega_q / E_{C_J}$, as opposed to $\omega_q / E_{C,\rm total}$ in the fluxonium, with $E_{C,\rm total}$ the charging energy of the combined junction and shunt capacitance. Since $E_{C,\rm total} \ll E_{C_J}$, this implies a correspondingly smaller loss rate in our device. This protection arises from our circuit topology that separates the roles of the small junction capacitances and the large shunt capacitance for which symmetry protection applies. Using the previously estimated $Q_\mathrm{cap}$, we predict $1/\Gamma^\mathrm{JJ}_\mathrm{charge}\approx 70~\mathrm{ms}$ at half-flux.

\zl{A further refinement of our model would be to include the readout mode. A detailed analysis of its interaction with our qubit \cite{Zhu2013} will be addressed in future work. More decay channels could be uncovered by including JJ array modes \cite{Masluk2012}, as well as accounting for quasiparticles \cite{pop2014}.}

\section{Pathway to high coherence}

\begin{figure}
    \centering
    \includegraphics[scale=1]{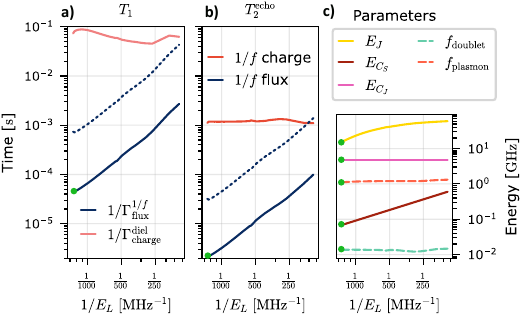}
    \caption{\zl{{\bf Flux noise mitigation.} (a) Simulated decay time (y-axis) versus inverse inductive energy (x-axis). The contribution from dielectric loss (red) remains largely unaffected, while the contribution from $1/f$ flux noise (blue) improves by two orders of magnitude. (b) Simulated coherence time (y-axis) versus inverse inductive energy (x-axis). The contribution from $1/f$ charge noise (red) remains largely unaffected, while the contribution from $1/f$ flux noise (blue) improves by two orders of magnitude. In (a,b), two values of flux noise levels are used: $A_\Phi= 5.25~\mu\Phi_0/\sqrt{\mathrm{Hz}}$ (blue solid lines) and an improved $A_\Phi= 1.4~\mu\Phi_0/\sqrt{\mathrm{Hz}}$ (blue dashed lines). Other circuit parameters are varied along with $E_L$, while the doublet and plasmon frequencies are kept constant, as reported in (c). Green dots indicate the parameters of the measured device.}}
    \label{fig:roadmap}
\end{figure}

\zl{The preceding sections identify $1/f$ flux noise as the dominant limitation of the present device. In this section we describe a parameter regime that reduces flux sensitivity by two orders of magnitude , without compromising charge protection. As with the flux qubit to Fluxonium to Blochnium evolution, we move towards lower values of $E_L$. For each value of $E_L$, we adapt $E_J$ and $E_{C_S}$ to satisfy three constraints. First, we fix the plasmon frequency, preventing it from crashing down to low frequencies, which would trigger thermally induced transitions and compromise the spectral gap of the qubit. Second, we fix the doublet frequency, preventing an increase of the charge dispersion. Third, we fix $E_{C_J}$, thereby preserving the high bound for dielectric loss in the small JJs (Eq.~\eqref{eq:JJdiel}). The resulting decay and coherence times, reported in Fig.~\ref{fig:roadmap}, are extracted from a 3-mode model (\ref{sec:t1} and \ref{sec:t2}). We find that moving towards $E_L/h = 150~\mathrm{MHz}$, for the flux noise amplitude measured in this experiment ($A_\Phi= 5.25~\mu\Phi_0/\sqrt{\mathrm{Hz}}$), we would reach a flux-limited $T_1=2.7~\mathrm{ms},~T_2^\mathrm{echo}=99~\mu\mathrm{s}$, and with state of the art noise levels ($A_\Phi= 1.4~\mu\Phi_0/\sqrt{\mathrm{Hz}}$ \cite{Yan2016}), we reach $T_1=43~\mathrm{ms},~T_2^\mathrm{echo}=1.6~\mathrm{ms}$. The decay and decoherence due to dielectric loss and charge noise respectively, are unaffected. Note that charge-noise-induced decoherence may be improved without compromising on flux-noise by lowering the doublet frequency while adapting the inductive energy. Remarkably, this shows that we can match the flux-induced coherence times of the Fluxonium, and at the same time, preserve the protection from dielectric loss. This evades the charge-flux trade-off predicted by Ref.~\cite{messelot2026} based on a single-mode analysis similar to Eq.~\eqref{eq:H1M}. In this newly uncovered regime, the Born-Oppenheimer gap that controls the single-mode reduction closes, and the full 3-mode model must be considered \cite{Roverchtheory}.}

\zl{We emphasize that the value of such a device does not rest solely on its prospects as a transmon or fluxonium replacement. A qubit engineered with a high, quantitatively predicted bound against one or more well-understood noise mechanisms is a powerful diagnostic tool: any coherence measured below that bound signals the presence of an unidentified loss channel. Historically, such qubits where dominant loss mechanism were deliberately reduced have been instrumental in uncovering new physics in superconducting circuits, e.g. quasiparticle-induced loss in fluxonium \cite{pop2014}. A $\cos(2\varphi)$ qubit combining high charge- and flux-noise bounds, even if control and readout remain challenging, would offer a similarly clean testbed for isolating remaining, currently unidentified decoherence channels in superconducting circuits.}

\section{Conclusion and outlook}
In conclusion, we implement a superconducting qubit encoded in states of opposite Cooper-pair parity. We operate in the soft transmon regime at a qubit frequency of 13.6 MHz while preserving coherent control and single shot readout. In this regime, our qubit is no longer limited by charge-related mechanisms, such as dielectric loss and charge noise, but rather by $1/f$ flux noise. Our measurements thus establish that Cooper-pair parity symmetry shields the qubit from the dielectric loss it was designed to suppress, leaving the interferometer’s flux sensitivity as the dominant limitation. \zl{Moving towards devices in the flux-resilient regime, combined} with reduced loop areas, gradiometric layouts, or protection via fluxoid quantization, \zl{is} expected to improve this limit. On the longer run, exciting perspectives could arise from novel materials that implement quartet tunneling without a flux loop, such as twisted $d-$wave superconductors \cite{Brosco2024flowermon}. 

More generally, our circuit combines key characteristics of a heavy fluxonium with low-lying transitions and high anharmonicity, and of a Cooper-pair box with quantized charge number and sensitivity to quasiparticle tunneling. Endowed with the ability to measure in a single shot both the quasiparticle parity and the Cooper-pair parity, our device, operating as a detector, could shed light on the puzzling mechanisms of the creation of quasiparticles and their energy distribution \cite{Catelani2011,Serniak2018}.

\section*{Author Contributions}
E.R. measured the device \zl{and discovered the flux-protected regime}, A.B. fabricated the sample and M.V. implemented the microwave and cryogenic setup and fabricated initial versions of the device. E.R, A.B. and M.V. designed the device. E.R, K.G. and Z.L. analyzed the data. K.G. developed the Hamiltonian fitting routine with W.C.S. and A.P. for guidance. B.D. provided theory support. P.C.I. and C.W. provided experimental support. E.R. and Z.L. wrote the manuscript with input from all authors. E.R, W.C.S and Z.L. conceived the experiment.

\section*{Acknowledgements}
We thank José Palomo, Aurélien Schmitt and Aurélie Pierret for providing nano-fabrication facilities, and Taha Bouwakdh for insightful numerical simulations. This work was supported by the QuantERA grant QuCOS, by ANR 19-QUAN-0006-04. This project has received funding from the European Research Council (ERC) under the European Union’s Horizon 2020 research and innovation program (grant agreement no.\ 851740). This work has been funded by the French grants ANR-22-PETQ-0003 and ANR-22-PETQ-0006 under the ``France 2030 Plan.''

\bibliographystyle{apsrev4-1}
\bibliography{biblio}
\begin{figure*}[ht]
    \centering
    \includegraphics[scale=0.8]{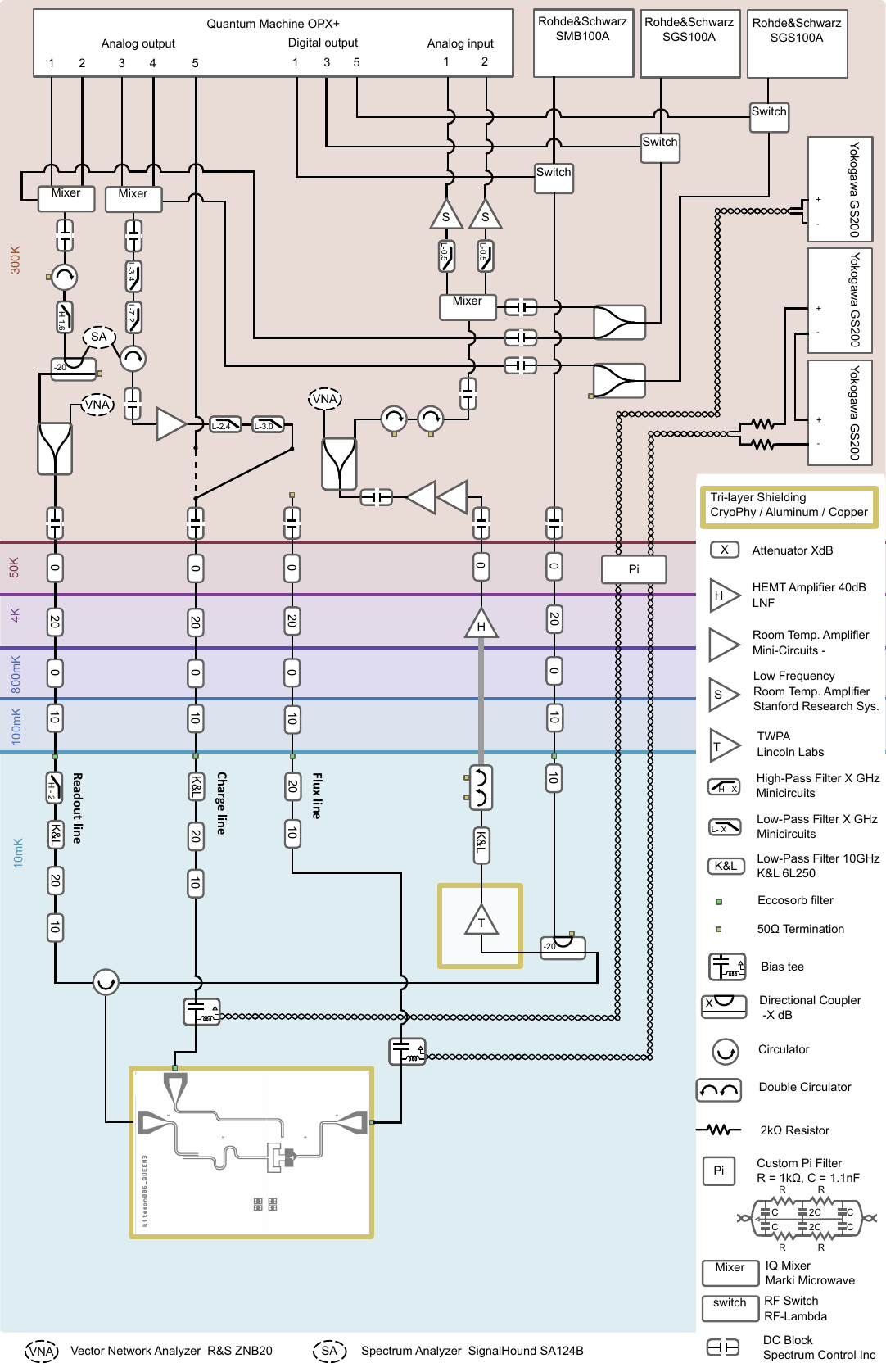}
    \caption{{\bf Wiring diagram of the experiment.} Straight lines denote coaxial cables, while the interleaving curves represent twisted pairs. The thick gray line represents a NbTi superconducting output line in the cryostat. The dashed lines at the input of the fridge represent the two wiring configurations used in the experiment, depending on the qubit drive frequency range.}
    \label{fig:cabling}
\end{figure*}
\clearpage
\setcounter{section}{0}
\renewcommand{\thesection}{Appendix \Alph{section}}
\renewcommand{\thesubsection}{\arabic{subsection}}

\section{Sample and setup}
\label{sec:samplesetup}
\subsection{Fabrication}
We fabricate an array of $3\times 3$ chips on a 2-inch wafer. From chip to chip, we vary the area of the small KITE junctions, and after dicing, we select the chip that is closest to the target parameters. In the following, we detail each step of the fabrication process.

\paragraph{Wafer preparation:}
The circuit is fabricated on a $280\ \mu \mathrm{m}$-thick wafer of 0001-oriented, double-side polished high resistivity ($\geq 20\ \mathrm{k}\Omega.\mathrm{cm}$ ) intrinsic silicon. The silicon wafer is initially cleaned in a 3:1 $\mathrm{H}_2\mathrm{S}\mathrm{O}_4$:$\mathrm{H}_2\mathrm{O}_2$ Piranha solution at room temperature for 5 minutes followed by a 5:1 buffered oxide etch for 5 minutes, after which it is loaded into a sputtering system. After one night of pumping, $150\ \mathrm{nm}$ of niobium are sputtered with argon plasma.

\paragraph{Circuit patterning:} We spin optical resist (S1805) and pattern the large features (on-chip bias lines and capacitor pads) using a laser writer. After development (MF319), we rinse in de-ionized water for $1$ minute, and etch the sample in SF6 with a $20\ \mathrm{s}$ over-etch. Finally, the sample is cleaned for 10 minutes in acetone at $50~\degree\mathrm{C}$ followed by $30\ \mathrm{s}$ stripping process in a reactive ion etching (RIE) machine.

\paragraph{Junction patterning: } Next, we remove the oxide regrowth on Nb and Si by applying another round of 5:1 buffered oxide etch for 5 minutes. The wafer is then rinsed in water, spun dry, and quickly covered with a bilayer of methacrylic acid/methyl methacrylate [MMA (8.5) MAA EL13] and poly(methylmethacrylate) (PMMA A3). The KITE circuit (cross junctions, junction arrays, and connecting leads) is patterned in a single e-beam lithography step with a 20-keV Raith electron-beam pattern generator. The development takes place in a 3:1 isopropyl alcohol (IPA)/water solution at $6~\degree\mathrm{C}$ for $90\ \mathrm{s}$, followed by $10\ \mathrm{s}$ in IPA.

\paragraph{Junction deposition: } The wafer is then loaded in an e-beam evaporator. We start with a thorough argon ion milling for 44 seconds at $\pm30 \degree $ angles. We then evaporate $30\ \mathrm{nm}$ and $100\ \mathrm{nm}$ of aluminum, at $\pm30\degree$ angles, separated by an oxidation step in $5 \ \mathrm{mbar}$ of pure oxygen for 5 minutes, for Josephson junctions with critical current density of $100\ \mathrm{A}/\mathrm{cm}^2$. The wafer is then inserted in N-Methyl-2-pyrrolidone (NMP) at 80~$\degree \mathrm{C}$ for liftoff until all excess aluminium is removed (typically 1.5 hours).

\paragraph{Junction characteristics: }
The Josephson junctions are all fabricated from Al/AlOx/Al in a single evaporation step, utilizing the Dolan bridge method. The e-beam base dose is set to $283~\mu\mathrm{C}/\mathrm{cm}^2$, with an acceleration voltage of $20~\mathrm{kV}$ and a lens aperture of $7.5~\mu\mathrm{m}$. Two types of junctions are fabricated. (i) Two small cross junctions are located in the KITE, with an area of $0.042~\mu\mathrm{m}^2$. They are patterned with a dose factor of $1.1$ and an undercut dose of $0.2$, resulting in an inductance per junction of $6~\mathrm{nH}$. (ii) A total of $300$ large array junctions are also located in the KITE loop. These junctions have an area of $0.48~\mu\mathrm{m}^2$, and are patterned with a dose factor of $1.0$ and an undercut dose of $0.1$, resulting in an inductance per junction of $0.65~\mathrm{nH}$. The small junction inductances turned out slightly too small compared to the target parameter regime. In order to reduce the $E_J/E_{C_J}$ ratio, the device was baked for 90 minutes at $130~\degree\mathrm{C}$, increasing the value of the inductances of the cross junctions and the arrays to $9.5~\mathrm{nH}$ and $0.86~\mathrm{nH}$, respectively.
\paragraph{Dicing: }
We apply a protective bilayer of poly(methyl methacrylate) (PMMA A6) and optical resist (S1805) and dice the wafer with an automatic dicing saw. The protective layers on the chips are subsequently removed in two steps with 4\%~KOH and acetone.

\subsection{Wiring}
The chip, which contains the device of interest, a larger-capacitance replica, and several test junctions, is mounted with PMMA and wire-bonded to the PCB of the sample holder denoted ``JAWS'' in Ref.~\cite{mariusthesis}. The sample holder is then mounted onto the base plate of a Bluefors LD250 dilution refrigirator, inside a trilayer shielding can. The complete wiring diagram of the experiment is shown in Fig.~\ref{fig:cabling}. The traveling-wave parametric amplifier (TWPA), provided by Lincoln Labs, is mounted in a separate can and provides around 15 dB of amplification away from its dispersive feature at 6 GHz when powered by the SMB100A RF source. The two SGS100A sources provide LO tones for the readout and qubit drives. All three sources are gated by RF switches controlled by the digital outputs of the OPX+, such that the LO signal is only let through at the designated times in experimental sequences.

The qubit is driven, for the plasmonic transitions, through an IQ mixer (X-Microwave XM-C9P7-0609C) with a 2-18 GHz RF bandwidth. Since the lowest-frequency transition we need to address with this mixer lies around 1 GHz, we compensate for the increased out-of-band insertion loss by adding an amplifier after the upconversion module; low-pass filtering both before and after the amplifier helps mitigate the spurious LO harmonics generated by the mixer.
For frequencies under 200 MHz (doublet transition), the qubit is driven directly from the IF output of the OPX+.
The downconversion module consists of another IQ mixer: both quadratures of the downconverted signal are routed to the OPX+ to perform dual demodulation. The two digitized signals are combined with a relative phase that maximizes SNR, thereby mimicking an image reject mixer.

All RF instruments are referenced to a
Stanford Research Systems FS725 Rubidium clock.
DC-current biasing of the flux line is carried out by two Yokogawa GS200 voltage sources in parallel with different range settings, simultaneously providing large offset compensations and a high degree of precision for biasing around the half-flux symmetry point. The third GS200 source provides DC charge biasing.
\section{Flux and charge bias calibration}
\label{sec:biascal}

\begin{figure}[h!]
    \centering
    \includegraphics[scale=1]{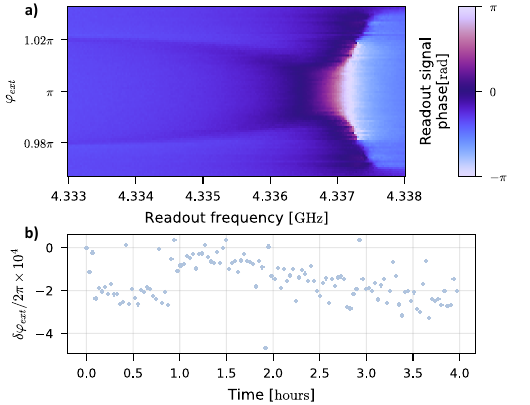}
    \caption{{\bf External flux bias calibration.} (a) Reflected signal phase (color) versus probe frequency (x-axis) and external flux (y-axis). The two features used for calibration are visible to the left of the main resonance frequency. (b) External flux drift (y-axis) versus time (x-axis) over a 4-hour measurement run. The scatter in the data (dots) is likely due to imprecisions in the fitting procedure rather than rapid fluctuations. From $t=1.5~\mathrm{hour}$ to $t=3.5~\mathrm{hours}$, we observe a downward trend of $\delta\phiext/2\pi\approx 0.2\times 10^{-3}$, corresponding to a drift of the order of $0.1~\mathrm{m}\Phi_0$ per hour.}
    \label{fig:fluxbias}
\end{figure}
The external biases $\phiext$ and $N_g$ provide tunable knobs to explore the qubit landscape, but also tend to drift due to environmental fluctuations. In order to guarantee the accuracy of the acquired data, we interleave calibrations of both flux and charge within our measurement sequences.

\subsection{Flux}

To compensate for external flux drifts, we use the sharp readout anticrossing features shown in Fig. \ref{fig:fig2}a as a flux reference.
Specifically, a pair of anticrossings lying around $\pm 0.01 \Phi_0$ from the symmetry point provides a good syndrome when probed away from the readout resonance.

The calibration process consists in probing the resonator at frequency $4.331~\mathrm{GHz}$ on two $5~\mathrm{m}\Phi_0$ spans around the expected values of $\Phi_0\times(0.5\pm 0.01)$. We fit the phase roll of each anticrossing versus DC voltage and record the corresponding central voltages $V_\pm$. The voltage corresponding to half a flux quantum is then defined as $V_\pi = (V_+ + V_-)/2$. Having previously measured the flux periodicity in units of voltage $V_{2\pi\mathrm{-period}}$, we express the flux drift over time $t$ as $\delta\phiext(t)/2\pi=(V_\pi(t)-V_\pi(0))/V_{2\pi\mathrm{-period}}$.
    
Note that this process can be performed independently of the value of the offset charge $N_g$, removing the need for cross-calibration of both biases; however, it requires a coarse prior knowledge of the half-flux symmetry point location so that the anticrossing features are contained in the spectroscopy window.

During data acquisition, this calibration sequence is performed every $\approx 100$ seconds. Fig \ref{fig:fluxbias}b shows that the flux drift timescale is on the order of $0.1~\mathrm{m}\Phi_0$ per hour.

\subsection{Charge}
\begin{figure}[h!]
    \centering
    \includegraphics[scale=1]{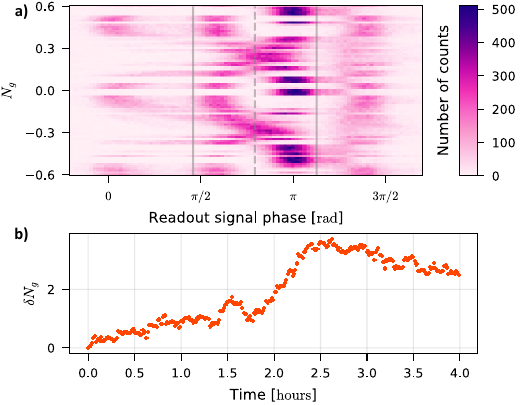}
    \caption{{\bf Offset charge bias calibration.} (a) Histograms of the measured $(I,Q)$ phase (x-axis) versus charge bias (y-axis). The full and dashed lines respectively denote the boundaries and symmetry axis of the calibration procedure detailed below.}
    \label{fig:chargebias}
\end{figure}
Offset charge calibration poses a more subtle challenge, owing to the coexistence of the two quasiparticle branches on the typical timescale of a calibration sequence. These interleaving branches lead to the usual eye pattern shown in Fig.~\ref{fig:chargebias}a, where the typical quasiparticle jump time is longer than the acquisition time of one horizontal slice, but shorter than that of the full measurement (see \ref{sec:quasiparticles}).

We collect the raw readout signal histograms as in Fig.~\ref{fig:fig2} to avoid averaging effects due to qubit transitions or quasiparticle jumps. Having isolated the region in $IQ$ plane enclosing $\ket{0-, \mathrm{even}}$ and $\ket{0\pm, \mathrm{odd}}$ (full grey lines in Fig.\ref{fig:chargebias}a), we numerically correct for quasiparticle jumps by reflecting the data along the dashed grey line. The distributions along the readout phase are then averaged for every $N_g$, and the resulting averaged signal is fit to a function of the form $\abs{\cos(2\pi(N_g-\delta N_g))}$, where the offset $\delta N_g$ is the only fit parameter. We then repeat the process on a narrower charge span to increase precision.
    

During data acquisition, this calibration sequence is
performed every $\approx30$ seconds. Fig \ref{fig:chargebias}b shows that charge drifts occur much faster than flux drifts, with a full period being spanned in $\approx1$ hour.
\section{Quasiparticle tunneling}
\label{sec:quasiparticles}

Quasiparticles carrying a $1e$ charge tunnel in and out of the qubit island on a timescale slower than that of Cooper-pair dynamics.
\begin{figure}[h!]
    \centering
    \includegraphics[scale=1]{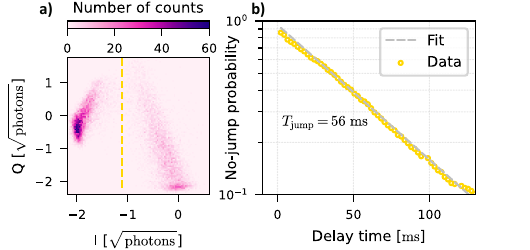}
    \caption{{\bf Quasiparticle jumps} (a)  Two-dimensional histograms of the readout $(I,Q)$ values at $N_g=0$ and $\phiext=\pi$, integrated over $200~\mu\mathrm{s}$. Two distributions appear, corresponding to even (left of the yellow dashed line) and odd (right) quasiparticle parity. (b) No-jump probability (y-axis) versus time between two single-shot measurements of quasiparticle parity (x-axis). The data follows an exponential law (grey dashed line), indicating a memory-less Poissonian process characterized by $T_\mathrm{jump} = 56~\mathrm{ms}$. }
    \label{fig:qpjumps}
\end{figure}
By integrating the readout signal for $200~\mu\mathrm{s}$, we average out the qubit dynamics shown in Fig.\ref{fig:fig2} while retaining distinct distributions for either quasiparticle manifold; Fig.\ref{fig:qpjumps}a shows the single-shot discrimination of quasiparticle parity through this measurement. Analyzing the time correlations of these parity measurements (Fig.\ref{fig:qpjumps}b) yields a typical jump time of $T_\mathrm{jump} = 56~\mathrm{ms}$. We then perform the quasiparticle parity measurement every few milliseconds during the sequences presented in the main text, and only keep the data corresponding to the even manifold. Data pertaining to the odd manifold (equivalently $N_g=0.5$) are shown in \ref{sec:oddmanifold}. 
\section{Characterization at zero flux}
\label{sec:zeroflux}
\begin{figure}[h!]
    \centering
    \includegraphics[scale=1]{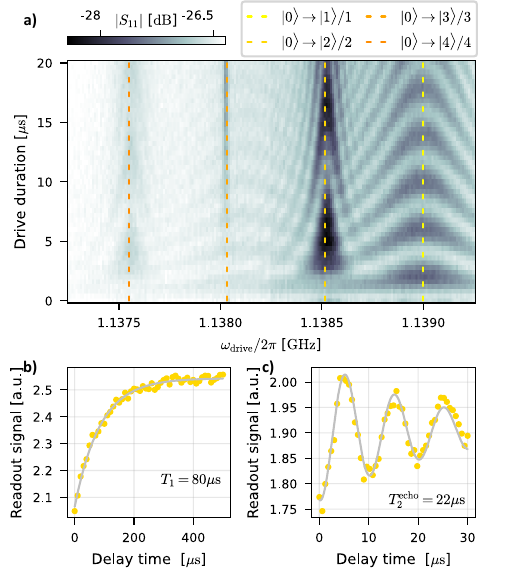}
    \caption{{\bf Zero-flux plasmon characterization} (a) Reflected readout signal amplitude versus qubit drive frequency (x-axis) and pulse duration (y-axis). The dashed lines show the resonant frequencies of the first few multi-photon plasmonic transitions. (b) Readout signal amplitude (y-axis) versus delay time (x-axis) after a $\pi$-pulse is applied to the first plasmonic transition. The data (dots) are fit to a decaying exponential (solid line) leading to an energy relaxation time $T_1 = 80~\mu\mathrm{s}$. (c) Readout signal amplitude (y-axis) versus total wait time (x-axis) between the two $\pi/2$ qubit pulses separated by the echo $\pi$ pulse. The data (dots) are fit to an exponentially decaying cosine function (solid line), leading to an echo decay time $T_2^\mathrm{echo} = 22~\mu\mathrm{s}$.}
    \label{fig:0flux}
\end{figure}
The qubit spectrum at $\phiext=0$ resembles that of a quasi-harmonic oscillator, featuring only single-well plasmonic excitations (unlike the fluxonium qubit at zero flux which still features secondary wells and a large spectral anharmonicity \cite{Ardati2024,Mencia2024}). The readout dispersive coupling being smaller than at the half-flux symmetry point, we do not have access to single-shot state discrimination and therefore plot the data in Fig.\ref{fig:0flux} in terms of the reflected signal magnitude $|S_{11}|$.

Driving the qubit through the charge line around the plasmonic transition frequency, we obtain canonical Rabi chevrons of the lowest-energy transition $\ket{0} \leftrightarrow \ket{1}$. Evenly-spaced chevron patterns are also visible to the left of the main one: these correspond to coherent multi-photon excitations of the higher plasmons, consistent with a 4th-order nonlinear Kerr oscillator. From the spacing between the symmetry axis of each pattern, we extract a $1~\mathrm{MHz}$ Kerr anharmonicity for the plasmonic ladder. This value is two orders of magnitude smaller than that of a typical transmon. For comparison, an equivalent transmon circuit with the same frequency and anharmonicity would have a Josephson energy of 162 GHz and a charging energy of 1 MHz \cite{Koch2007}. Here the junction's nonlinearity is diluted by the large series inductance.

\section{Odd quasiparticle parity manifold}
\label{sec:oddmanifold}

\begin{figure}[h]
    \centering
    \includegraphics[scale=1]{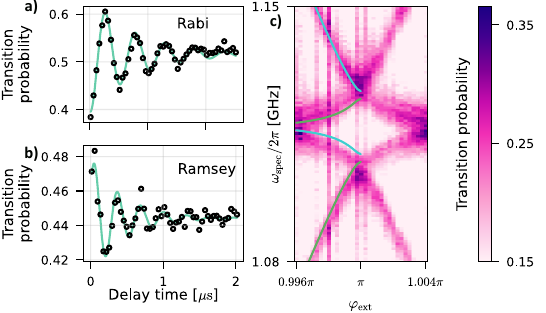}
    \caption{{\bf Characterization of the odd manifold}. (a) Rabi oscillations of the $(\ket{0+,\mathrm{ odd}}, \ket{0-,\mathrm{ odd}})$ doublet, decaying with a characteristic time $T_2^\mathrm{Rabi} = 800~\mathrm{ns}$. (b) Ramsey spectroscopy of the doublet, from which we extract $\omega_{q,\mathrm{odd}}= 2\pi \times 2.7~\mathrm{MHz}$ and $T_2= 440~\mathrm{ns}$. The signal oscillations arise from the combined digital and physical detunings of the $\pi/2$ pulses with respect to the bare qubit frequency. (c) Two-tone spectroscopy versus external flux around $\phiext=\pi$ and the first plasmonic transition, with solid lines showing the transition frequencies from $\ket{0+,\mathrm{ odd}}$ (blue) and $\ket{0-,\mathrm{ odd}}$ (green) computed with the 3-mode Hamiltonian \eqref{eq:H3M}. }
    \label{fig:odd}
\end{figure}

We characterize our circuit at odd quasiparticle number, or equivalently at $N_g=0.5$. We set $\phiext=\pi$, and locate the lowest doublet transition through two-tone spectroscopy. We perform Rabi oscillations (Fig.\ref{fig:odd}a) and further refine the measurement of our doublet transition frequency through a Ramsey sequence (Fig.\ref{fig:odd}b). We find $\omega_{q,\mathrm{odd}}=2\pi\times 2.7~\mathrm{MHz}$. In comparison to $N_g=0$,  the Rabi and Ramsey decay times are lower ($800~\mathrm{ns}$ and $400~\mathrm{ns}$ respectively), which is attributed to the even sharper flux dispersion.

The transition frequency sets the value of the loop asymmetry (Fig~\ref{fig:fig1}f), and Fig.\ref{fig:odd}c shows that the lumped circuit model is in good agreement with spectroscopic data for the odd quasiparticle parity manifold as well (here we show the transition probability out of the doublet without discriminating on the initial state within the doublet).

\section{Lumped circuit Hamiltonian}
\label{sec:fit}

\begin{figure}
    \centering
    \includegraphics[scale=1]{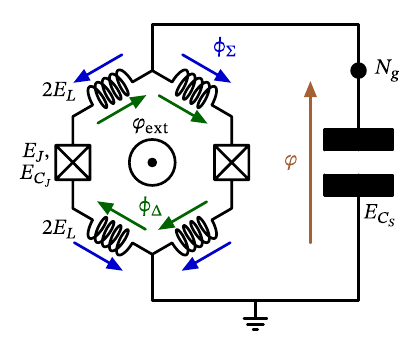}
    \caption{Circuit diagram of the KITE transmon, with colored arrows denoting the phase variables of the modes entering Hamiltonian \eqref{eq:H3M}.$\phi_\Sigma$ and $\phi_\Delta$ are respectively defined as the half-sum and half-difference of the phase drops along each arm's inductors.}
    \label{fig:circuitarrows}
\end{figure}
We derive the Hamiltonian of our circuit (Fig.~\ref{fig:circuitarrows}) by following the standard circuit quantization recipe \cite{Smith2019}, yielding:
\begin{eqnarray} 
\label{eq:H3M}
&&H^\text{3-modes}_\mathrm{circuit} = 4E_{C_S} (\hat N + \hat n_{\Sigma} - N_g)^2 + 2E_{C_J}(\hat n_{\Sigma}^2 + \hat n_{\Delta}^2)\notag
 \\
&&+  E_L \big(\hat \phi_{\Sigma}^2 + (\hat \phi_{\Delta}-\frac{\phiext}{2})^2\big) - 2E_J \cos{(\hat \phi_{\Delta})} \cos{( \hat \varphi - \hat \phi_{\Sigma}})\notag\\
&&+H_\mathrm{asym} \;.
\end{eqnarray}
This Hamiltonian describes the interaction of three modes. A compact and heavy mode represented by conjugate variables $(\hat \varphi, \hat N )$, and two light modes represented by conjugate variables $(\hat \phi_{\Sigma}, \hat n_{\Sigma} )$, $(\hat \phi_{\Delta}, \hat n_{\Delta} )$. The latter are the symmetric and antisymmetric modes internal to the KITE loop. In the regime where $E_{C_S}\ll E_{C_J}$, these light modes shape an effective $\cos(2\varphi)$ potential for the heavy mode that in turn enters Eq.~\ref{eq:H1M}. A rigorous mathematical derivation of the mapping from Eq.~\ref{eq:H3M} to Eq.~\ref{eq:H1M} is undergoing \cite{Roverchtheory}.

Asymmetry in the KITE loop is encompassed in the term
\begin{align}
    H_\mathrm{asym} = &- \frac{\Delta C_J}{C_J}\times 4E_{C_J}{\hat n_{\Sigma}} {\hat n_{\Delta}} -\frac{\Delta L}{L} E_L {\hat \phi_{\Sigma}} { (\hat\phi_{\Delta}- \frac{\phiext}{2})} \notag \\
    &+2\Delta E_J \sin{({\hat \phi_{\Delta}})} \sin{( {\hat \varphi} - {\hat \phi_{\Sigma}})} \;,
    \label{Hcircuit}
\end{align}
where 
\begin{align*}
    \Delta C_J &= (C_J^\mathrm{left} -  C_J^\mathrm{right})/2 \\
    \Delta L &= (L^\mathrm{left} -  L^\mathrm{right})/2 \\
    \Delta E_J &= (E_J^\mathrm{left} -  E_J^\mathrm{right})/2 \ ;
\end{align*}
and $E_{C_J}$, $E_L$, $E_J$ are defined as the mean values between the components' nominal energies on each arm.

The superinductance on each arm is made of 150 large junctions. The relative asymmetry $\Delta L/L$, that is expected to diminish with increased junction number, is therefore neglected. We further assume that asymmetry mainly arises from small variations in the area of the two small junctions, but keeping constant the plasma frequency $\omega_\mathrm{plasma} = \sqrt{8E_J E_{C_J}}/\hbar$. We are left with a single fit parameter $\varepsilon$ accounting for KITE asymmetry: $\varepsilon = \displaystyle \frac{\Delta E_J}{E_J} = \frac{\Delta C_J}{C_J}$. 

The two-tone spectroscopy data in Fig.~\ref{fig:fig3} and Fig.~\ref{fig:odd} are subject to systematic shifts due to the AC Stark effect. We therefore extract the unbiased transition frequencies at multiple external flux values and $N_g=0, 0.5$ from Ramsey measurements. The resulting data are fit using a gradient-descent procedure over the five parameters listed in Table~\ref{table1}.


\section{Photon number calibration}
\label{sec:photonnumber}

\begin{figure}
    \centering
    \includegraphics[scale=1]{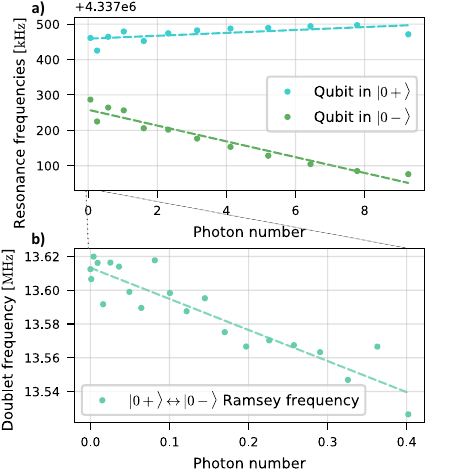}
    \caption{{\bf Dispersive shift and readout Stark shift of the doublet} (a) Readout resonator frequency (y-axis) versus photon number in the resonator (x-axis) conditioned on the qubit being in either state $\ket{0+}$ or state $\ket{0-}$. The dashed lines fit to the model of Eq.\eqref{readout_kerr}. (b) Doublet transition frequency obtained from Ramsey spectroscopy (y-axis) versus photon number in the resonator (x-axis). The dashed line corresponds to Eq.\eqref{stark} and sets the scale of the x-axis. Although the x-axes are plotted in photon number units for clarity using the calibration results, these measurements were initially performed in units of applied readout power; photon number calibration then consists in obtaining the conversion factor between these units.}
    \label{fig:photonnumber}
\end{figure}
We use the $(\ket{0+}, \ket{0-})$ doublet at $\pi$ to calibrate the steady-state photon number in the resonator during readout. This calibration is done in two steps: first, we measure the dispersive shift of the doublet by probing the resonance frequency of the readout conditioned on the qubit state (Fig.\ref{fig:photonnumber}a). The readout power is swept in order to capture the qubit-state-dependent Kerr effect, as featured in this effective model \cite{Zhu2013}:
\begin{align}
    H_\mathrm{readout}/\hbar = \omega_0 \hat a^\dagger \hat a &+(\chi_+\hat a^\dagger \hat a  - \frac{K_+}{2} {{\hat a}^{\dagger 2}} {\hat a}^2) \ket{0+}\bra{0+} \notag \\
    & +(\chi_-\hat a^\dagger \hat a  - \frac{K_-}{2} {{\hat a}^{\dagger 2}} {\hat a}^2) \ket{0-}\bra{0-}
    \label{readout_kerr}
\end{align}
where $\omega_0$ is the bare resonator frequency, $\hat a$ is the corresponding annihilation operator, and $\chi_\pm$ and $K_\pm$ denote the second- and fourth-order nonlinearities, respectively, for each state of the doublet.
The intercepts of the linear fits of Fig.\ref{fig:photonnumber}a yield the doublet dispersive shift $$\chi/2\pi =(\chi_- - \chi_+)/2\pi= -201~\mathrm{kHz}\ .$$
Note that unlike $K_+$ and $K_-$, the fit for $\chi$ does not require foreknowledge of the photon number.

The second step is then to quantify the AC Stark shift $\Delta\omega_q$ of the doublet by measuring its transition frequency versus readout power (Fig.\ref{fig:photonnumber}b). AC Stark shift is given by:
\begin{equation}
    \Delta \omega_q = \bar{n} \chi\;,
    \label{stark}
\end{equation}
where $\bar n$ is the number of photons in the resonator. To perform this measurement, we remain at low readout power so that 4th-order terms in Eq.\eqref{readout_kerr} are negligible. Knowing the value of $\chi$, the linear fit in Fig.\ref{fig:photonnumber}b finally gives the conversion between applied readout power and photon number in the resonator. The slopes of the linear fits of Fig.\ref{fig:photonnumber}a then give $K_+/2\pi = -4~\mathrm{kHz}$, $K_-/2\pi = 22~\mathrm{kHz}$.
\section{Qubit relaxation}
\label{sec:t1}

\begin{figure*}[t]
    \centering
    \includegraphics[scale=1]{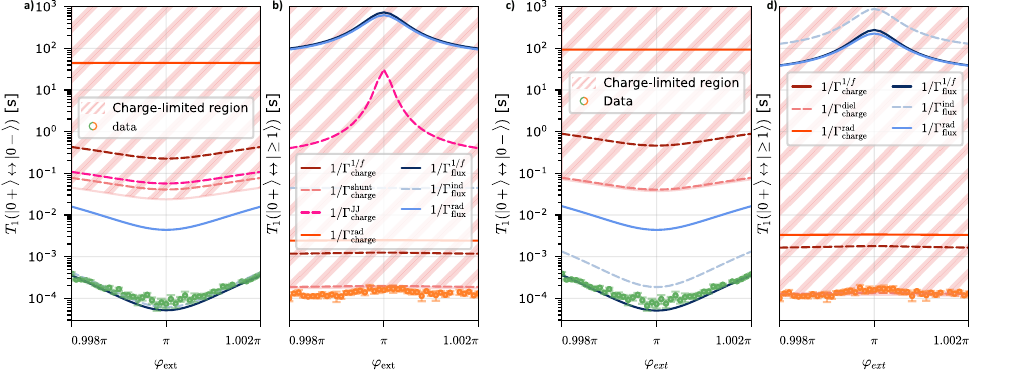}
    \caption{{\bf Overview of decay channels within the 3-mode and 1-mode models}.(a,b) Decay time (y-axis) of the doublet (a) and the plasmon (b) versus external flux (x-axis). The data (circles) are compared to calculated contributions from channels listed in the appendix within the 3-mode lumped circuit model. Solid lines correspond to values with no free fit parameters, while dashed lines represent lower-bound estimates. (c,d) Same as (a,b) with decay contributions computed from the single-mode model. }
    \label{t1_full}
\end{figure*}
Qubit relaxation is modeled using the Fermi golden rule \cite{Clerk2010}. We assume linear coupling of the system to its external environment through a Hamiltonian 
\begin{equation}
    \hat H_c = \hbar g \times \hat \Lambda_\mathrm{env} \times \hat{A} \ ,
\end{equation}
where $g$ is the coupling strength, $\hat{A}$ is a circuit operator and $\hat\Lambda_\mathrm{env}$ is an operator of the dissipative bath described by noise spectral density $S_\Lambda(\omega)$. The one-way decay rate from eigenstate $\ket{i}$ of eigenenergy $E_i$ to eigenstate $\ket{j}$ of eigenenergy $E_j$ is given by \cite{kyrylothesis}:
\begin{equation}
    \Gamma_{i\rightarrow j} = g^2 |\bra{i}\hat A\ket{j}|^2 S_\Lambda(\omega_{ji})\;,
    \label{eq:gamma_oneway}
\end{equation}
where $\omega_{ji}=(E_i-E_j)/\hbar$. In the case of a two-level system (TLS), decay is exponential with a time constant $T_1 = 1/\Gamma_{ij}$, where
\begin{equation*}
    \Gamma_{ij} = \Gamma_{i\rightarrow j} + \Gamma_{j\rightarrow i}\;.
\end{equation*}

We will now recall an identity that will be useful to compute the matrix element entering Eq.~\eqref{eq:gamma_oneway}. Considering an operator $\hat{A}$ and a pair of eigenstates $(\ket{j},\ket{k})$ of circuit Hamiltonian $\hat H$, the following holds
\begin{equation*}
    \bra{k}[\hat H, \hat A] \ket{j} = \hbar \omega_{jk} \bra{k}\hat A \ket{j}\;.
\end{equation*}
For a fluxonium with charging energy $E_\mathrm{C,total}$, this leads to the well-known expression linking the phase $\hat \phi$ and charge $\hat n$ matrix elements: $\displaystyle \bra{k}\hat n \ket{j} = \frac{\hbar \omega_{jk} }{8 i E_\mathrm{C,total}}\bra{k}\hat \phi \ket{j}$ \cite{Zhang2021}. With the 3-mode Hamiltonian of eq. \eqref{Hcircuit}, we obtain a similar relation for the $ \phi_\Delta$ mode,
\begin{equation}
    \hbar \omega_{jk} \bra{k}\hat \phi_\Delta \ket{j} = 4iE_{C_J} \bra{k}\hat n_\Delta \ket{j} -  4i \varepsilon E_{C_J} \bra{k}\hat n_\Sigma \ket{j}\;.
\label{matelt_delta}
\end{equation}
\subsection{Relaxation channels}

\subsubsection{Dielectric loss}
Dielectric loss in a capacitive element can be modeled as quantum Johnson-Nyquist noise from the real part of the element's impedance. The associated voltage spectral density reads \cite{Schoelkopf2003}:
\begin{equation}
    S^\mathrm{diel}_{VV}(\omega) = \hbar \omega \times \frac{1}{C|\omega| Q_\mathrm{cap}} \Big(1+ \coth(\frac{\hbar \omega}{2k_B T}) \Big) \ ,
\end{equation}
where $C$ is the capacitance of the element and $Q_\mathrm{cap}$ is the frequency-independent dielectric quality factor.
Summing the positive- and negative-frequency contributions gives the usual form

\begin{equation}
    S^\mathrm{diel}_{VV}(\omega) +S^\mathrm{diel}_{VV}(-\omega) = \frac{2\hbar }{CQ_\mathrm{cap}} \coth(\frac{\hbar |\omega|}{2k_B T}) \ ,
\end{equation}
from which we obtain Eq.\eqref{eq:diel} in the main text for the single-mode model.

On the other hand, applying these expressions to the various capacitors in the lumped circuit model gives :
\begin{equation}
    \Gamma_\mathrm{charge}^\mathrm{shunt}\big|_\text{3-modes} = \frac{16E_{C_S}}{\hbar Q_\mathrm{cap}}\coth(\frac{\hbar \abs{\omega_{ij}}}{2k_B T}) \abs{\bra{i}\hat N+\hat n_\Sigma \ket{j}}^2
\end{equation}
for the island capacitor; this expression vanishes for the doublet in the limit of no loop asymmetry.

For the doublet transition with a symmetric loop, 
the total dielectric loss rate associated to the small junction capacitors reads
\begin{align}
    \Gamma_\mathrm{charge}^\mathrm{JJ}\big|_\text{3-modes} &= \frac{8E_{C_J}}{\hbar Q_\mathrm{cap}}\coth(\frac{\hbar \omega_q}{2k_B T}
    ) \abs{\bra{0+}\hat n_\Delta \ket{0-}}^2 \notag \\
    &= \frac{\hbar\omega_q^2}{2E_{C_J} Q_\mathrm{cap}}\coth(\frac{\hbar\omega_q}{2k_B T})\abs{\bra{0+}\hat \phi_\Delta \ket{0-}}^2
\end{align}
where the last equality uses Eq.~\eqref{matelt_delta}.
\subsubsection{Inductive loss} Inductive loss, believed to arise from quasiparticle tunneling within the superinductances \cite{pop2014}, is similarly modeled using the real part of the inductors' admittance:
\begin{equation}
    S_{II}^\mathrm{ind}(\omega) +S_{II}^\mathrm{ind}(-\omega) = \frac{2\hbar }{LQ_\mathrm{ind}} \coth (\frac{\hbar |\omega|}{2k_B T}) \ ,
\end{equation}
where $Q_\mathrm{ind}$ is the frequency-independent inductive quality factor.

Summing contributions of the two arrays of the KITE loop (each with inductance $L$), we get
\begin{align}
    \Gamma_\mathrm{flux}^\mathrm{ind}\big|_\text{3-modes} = \frac{4E_{L}}{\hbar Q_\mathrm{ind}}\coth \left(\frac{\hbar |\omega_{ij}|}{2k_B T}\right) \Big(&\abs{\bra{i}\hat \phi_\Sigma \ket{j}}^2\\
    &+ \abs{\bra{i}\hat \phi_\Delta \ket{j}}^2 \Big)\notag
\end{align}

To obtain a similar estimate within the single-mode model without making assumptions about the microscopic origin of this loss channel, we apply the same reasoning to the flux-sensitive term in Hamiltonian \eqref{eq:H1M} :
\begin{equation}
    \Gamma_\mathrm{flux}^\mathrm{ind}\big|_\text{1-mode} = \frac{2\mathcal{E}_{J\phi}}{\hbar Q_\mathrm{ind}}\coth \left(\frac{\hbar |\omega_{ij}|}{2k_B T}\right) \abs{\bra{i}\sin \hat \varphi \ket{j}}^2 
\end{equation}
\subsubsection{$1/f$ flux noise} Flux noise arises from coupling of the inductive degrees of freedom to magnetic moments in the loop. Unlike the previous contributions, this noise spectral density shows no dependence on sample temperature:
\begin{equation}
    S_{\Phi\Phi}(\omega) = \frac{A_\Phi^2(2\pi \times1~\mathrm{Hz})}{|\omega|}
    \label{sphi}
\end{equation}
where $A_\Phi$ is the flux noise amplitude in units of magnetic flux per $\sqrt{\mathrm{Hz}}$ \cite{Yan2016} given in the main text. Counting both the $+\omega$ and $-\omega$ contributions, we obtain eq.\eqref{eq:1of} in the main text for the single-mode model. In the case of the 3-mode model, the equation becomes:
\begin{equation}
    \Gamma^{1/f}_\mathrm{flux}\big|_\text{3-modes}= \frac{8 \pi^2E_L^2}{\hbar^2 \Phi_0^2} \frac{A_\Phi^2(2\pi \times1~\mathrm{Hz})}{|\omega_{ij}|}\abs{\bra{i}\hat \phi_\Delta \ket{j}}^2 \;.
\end{equation}
The reported value of $A_\Phi=5.6~\mu \Phi_0 / \sqrt{\mathrm{Hz}}$ is obtained from the echo decoherence measurements (Fig.\ref{fig:fig5}d,e).

\subsubsection{$1/f$ charge noise}
Similarly, charge noise is assumed to follow a noise spectral density
\begin{equation}
    S_{QQ}(\omega) = \frac{A_Q^2(2\pi \times1~\mathrm{Hz})}{|\omega|} \ ,
    \label{Sqq}
\end{equation}
where $A_Q$ is the charge noise amplitude in units of electric charge per $\sqrt{\mathrm{Hz}}$.
The associated decay rate is written as
\begin{equation}
    \Gamma^{1/f}_\mathrm{charge}\big|_\text{1-mode}= \Big(\frac{8\mathcal{E}_C}{\hbar e} \Big)^2\frac{A_Q^2}{2}\frac{(2\pi \times1~\mathrm{Hz})}{|\omega_{ij}|}\abs{\bra{i}\mathcal{\hat N} \ket{j}}^2 \;
\end{equation}
within the single-mode model and
\begin{align}
    \Gamma^{1/f}_\mathrm{charge}\big|_\text{3-modes}= \Big(\frac{8E_{C_S}}{\hbar e} \Big)^2\frac{A_Q^2}{2}&\frac{(2\pi \times1~\mathrm{Hz})}{|\omega_{ij}|}\\
    &\times\abs{\bra{i}\hat N + \hat n_\Sigma \ket{j}}^2\notag \;
\end{align}
for the lumped circuit model.

We use a conservative estimate of $A_Q \leq 2 \times 10^{-3}e/\sqrt{\mathrm{Hz}}$ obtained from the echo decoherence versus offset charge (Fig.\ref{fig:fig5}f,g).
\subsubsection{Coupling to bias lines}
Decay also occurs from radiative loss through either of the bias lines, themselves coupled to a $R= 50~\Omega$ transmission line.

For the charge line, the coupling parameter is the gate capacitance $C_g = 0.16~\mathrm{fF}$, yielding
\begin{align}
    \Gamma^\mathrm{rad}_\mathrm{charge}\big|_\text{1-mode}= 4\pi \omega_{ij} \frac{R}{R_Q} \Big(\frac{C_g}{C} \Big)^2&\coth \left(\frac{\hbar |\omega_{ij}|}{2k_B T}\right) \\
    &\times \abs{\bra{i}\mathcal{\hat N} \ket{j}}^2\notag \;,
\end{align}
within the single-mode model and
\begin{align}
    \Gamma^\mathrm{rad}_\mathrm{charge}\big|_\text{3-modes}= 4\pi \omega_{ij} \frac{R}{R_Q} \Big(\frac{C_g}{C_S} \Big)^2&\coth \left(\frac{\hbar |\omega_{ij}|}{2k_B T}\right) \\
    &\times\abs{\bra{i}\hat N+\hat n_\Sigma \ket{j}}^2\notag \;
\end{align}
for the lumped circuit model, where $R_Q = \frac{h}{(2e)^2}$ is the superconducting resistance quantum and $C =  \frac{e^2}{2 \mathcal{E}_C}$.

The flux line is coupled through mutual inductance $M=2.1~\mathrm{pH}$, hence
\begin{equation}
    \Gamma^\mathrm{rad}_\mathrm{flux}\big|_\text{1-mode}= \frac{\omega_{ij}}{\pi}\frac{R_Q}{R} \Big(\frac{M}{L_\phi} \Big)^2\coth \left(\frac{\hbar |\omega_{ij}|}{2k_B T}\right)\abs{\bra{i}\sin \hat \varphi \ket{j}}^2 \;
\end{equation}
within the single-mode model, with $L_\phi = \frac{\Phi_0^2}{(2\pi)^2 \mathcal{E}_{J\phi}}$, and

\begin{equation}
    \Gamma^\mathrm{rad}_\mathrm{flux}\big|_\text{3-modes}= \frac{\omega_{ij}}{\pi}\frac{R_Q}{R} \Big(\frac{M}{L} \Big)^2\coth \left(\frac{\hbar |\omega_{ij}|}{2k_B T}\right)\abs{\bra{i}\hat \phi_\Delta \ket{j}}^2 \;
\end{equation}
for the lumped circuit model.
\subsection{Estimation of Transition Rates}

We distinguish in single shot three measurement outcomes: $\ket{0+}$, $\ket{0-}$, and a combined plasmon-excited outcome denoted $\ket{\geq 1}$.
This grouping is imposed by the measurement sequence, which does not resolve $\ket{1-}$ from $\ket{1+}$ or from higher plasmonic excitations with index $\geq 2$.
Since the thermal occupation of the latter states is negligible, the dynamics can be described within a three-level model formed by $\ket{0+}$, $\ket{0-}$, and $\ket{\geq 1}$.
We describe the dynamics with the vector of occupation probabilities $\mathbf{P}^T = \begin{pmatrix} P_{0+} \;\; P_{0-} \;\; P_{\geq 1}\end{pmatrix}$.

Assuming time-homogeneous decoherence, the state populations obey the master equation
\begin{equation*}
    \frac{d}{dt} \mathbf{P}
    =
    \mathbf{Q} \, \mathbf{P}
\end{equation*}
where the off-diagonal elements of the generator matrix are the transition rates, $\mathbf{Q}_{ij}=\Gamma_{j\rightarrow i}$ for $i\neq j$. Probability conservation, $P_{0+}+P_{0-}+P_{\geq 1}=1$, then fixes the diagonal elements such that each column of $\mathbf{Q}$ sums to zero.

To estimate these rates, we repeatedly measure the qubit while varying the waiting time between successive measurements; an example of such a sequence is shown in Fig.~\ref{fig:fig2}d. This procedure yields the statistics of transitions between the three outcomes as a function of the waiting time. As in the two-level case, the dependence is exponential in time, but is now governed by the matrix exponential $e^{\mathbf{Q}t}$, where $t$ is the time between two measurements. We fit the experimentally extracted transition statistics to this form, taking as free parameters the readout fidelities and three transition rates: $\Gamma_{0-\rightarrow 0+}$, $\Gamma_{0+\rightarrow 1\pm}$, and $\Gamma_{1\pm\rightarrow 0+}$, assuming symmetry with respect to parity. The resulting rates are used to obtain the lifetimes shown in Fig.~\ref{fig:fig5}b,c.

To estimate the statistical uncertainty on the extracted rates, we repeat the measurement and the reconstruction of the transition statistics 100 times. We then use a bootstrap procedure to resample these batches and repeat the fitting, thereby obtaining the error bars shown in Fig.~\ref{fig:fig5}b,c.

\section{Qubit dephasing}
\label{sec:t2}

We employ an echo sequence to probe the noise spectral densities of dephasing channels ($T_1$ contributions to decoherence are negligible for this device).
\subsection{Flux noise}
For first-order flux noise away from the sweet spot, the decay envelope $f$ at time $\tau$ is given by \cite{bylander2011}:
\begin{equation}
    f(\tau) = \exp \bigg(-\tau^2 \Big(\frac{\partial\omega_{q}}{\partial \phiext} \Big)^2 \int \mathrm{d}\omega \big(\frac{2\pi}{\Phi_0}\big)^2S_{\Phi\Phi}(\omega) g_1(\omega, \tau) \bigg)\ ,
    \label{echo_envelope}
\end{equation}
where $\omega_q$ is the angular frequency of the considered transition, $S_{\Phi\Phi}(\omega)$ is the noise spectral density defined in Eq.~\eqref{sphi} and $g_1(\omega, \tau)$ is a kernel function specific to the echo sequence.
In this experiment, we interleave one $\pi$-pulse of duration $\tau_\pi = 300 \mathrm{ns}$ in the middle of the waiting time, which leads to the expression:
\begin{equation}
    g_1(\omega,\tau) = \frac{1}{(\omega\tau)^2} \bigg| 1 + e^{i\omega\tau}-2 e^{i\omega\tau/2} \cos\frac{\omega \tau_\pi}{2}\bigg|^2 \underset{\omega \rightarrow 0}{\sim} \Big(\frac{\omega \tau}{4}\Big)^2
\end{equation}
Unlike a Ramsey sequence, this echo sequence features a vanishing kernel function at low frequencies, which will regularize the $1/f$ flux noise spectral density. This allows us to perform the analysis without the need for an infrared cutoff.

The envelope function we obtain is then fit to a Gaussian $f(\tau) \simeq \exp(-(\Gamma_\phi \tau)^2)$ \cite{bylander2011}; $1/\Gamma_\phi$ is the echo decay time associated to 1st-order flux noise.

However, this analysis does not take into account 2nd-order flux noise at the sweet spot, or any other flux-independent dephasing processes. To account for these processes near the sweet spot, we introduce an exponential flux-independent decay rate $\Gamma_\nu$; the total echo decay time $T_2^\mathrm{echo}$ is then defined as \cite{Zhang2021}:
\begin{equation}
    T_2^\mathrm{echo} =  \frac{\sqrt{\Gamma_\nu^2 + 4 \Gamma_\phi^2}-\Gamma_\nu}{2 \Gamma_\phi^2}
\label{t2echo}
\end{equation}

The data points in Fig.\ref{fig:fig5}d-e are obtained by fitting the time traces to an envelope $\propto \exp (-\Gamma_\nu^\mathrm{exp} t -(\Gamma_\phi^\mathrm{exp} t)^2)$, and the reported values of $T_2^\mathrm{echo}$ are computed through eq.\eqref{t2echo}.
In order to avoid overfitting, $\Gamma_\nu^\mathrm{exp}$ is kept fixed to the value of the decay rate obtained at $\phiext=\pi$, so that each trace only has one decay fit parameter.

The reported value of $A_\Phi=5.6~\mu \Phi_0 / \sqrt{\mathrm{Hz}}$ is obtained by fitting the values of $\Gamma_\phi^\mathrm{exp}$ to Eq.~\eqref{echo_envelope}.

\zl{The echo dephasing rate caused by flux noise at the sweet spot scales with $A_\Phi^2$ and the flux curvature of the qubit frequency $\displaystyle \frac{\partial^2 \omega_q}{\partial \phiext^2}$ \cite{Smith2019}.
Starting from the measured coherence at the sweet spot (assumed to be caused by flux noise via the flux calibration uncertainty), we use these scalings to estimate the values reported in Fig.~\ref{fig:roadmap}b.}

\subsection{Charge noise}
At first order, the charge noise dephasing envelope for the echo sequence is given by:
\begin{equation}
    f(\tau) = \exp \bigg(-\tau^2 \Big(\frac{\partial\omega_{q}}{\partial N_g} \Big)^2 \int \mathrm{d}\omega \frac{1}{(2e)^2} S_{QQ}(\omega) g_1(\omega, \tau) \bigg)\ ,
\end{equation}
The reported bound of $A_Q \leq \mathrm{} 2 \times 10^{-3}e / \sqrt{\mathrm{Hz}}$ is obtained by fitting the data in Fig.~\ref{fig:fig5}f,g to Eqs.~\eqref{t2echo} and \eqref{echo_envelope}.

\end{document}